\documentclass[a4paper,11pt]{article}

\usepackage{jcappub}

\usepackage{gensymb}

\usepackage[utf8]{inputenc}
\usepackage[english]{babel}
\usepackage[T1]{fontenc}
\usepackage{amsmath}
\usepackage{amsfonts}
\usepackage{hyperref}
\usepackage{amssymb}
\usepackage{physics}
\usepackage{slashed}
\usepackage[makeroom]{cancel}

\begin{document}

\title{Extreme Mass Ratio Inspirals \\ with Scalar Hair}

\author[a]{Adrien Kuntz,}
\author[b]{Riccardo Penco}
\author[a]{and Federico Piazza}

\emailAdd{kuntz@cpt.univ-mrs.fr}
\emailAdd{piazza@cpt.univ-mrs.fr}
\emailAdd{rpenco@cmu.edu}
\affiliation[a]{Aix Marseille Univ, Universit\'{e} de Toulon, CNRS, CPT, Marseille, France}
\affiliation[b]{McWilliams Center for Cosmology, Department of Physics, Carnegie Mellon University, Pittsburgh, PA 15213, USA}

\date{\today}

\abstract{Stellar mass objects orbiting around supermassive black holes are primary targets for future gravitational wave detectors like LISA.  However, in theories beyond general relativity, the corresponding waveform templates are still relatively poorly known. 
We propose a universal description for these systems which applies to any black hole with a non trivial scalar profile, or scalar \emph{hair}.
To this aim, we use the effective field theory recently introduced by Franciolini \emph{et al.} to write the most general action for the perturbations of a spherically symmetric solution up to some given order in derivatives and/or number of fields. At any post-Newtonian order,  the background metric and the relevant operators can be encoded in a limited number of parameters
 which are readily calculated in some given scalar tensor model, as we show with a couple of examples. In terms of such parameters, we obtain an analytic expression for the dissipated power in the odd sector by solving perturbatively the Regge-Wheeler equation in the presence of a point-particle source.}

\maketitle

\newpage

\section{Introduction}

The direct detection of gravitational waves (GWs) offers a new and exciting opportunity for testing General Relativity (GR). In particular, the future spatial interferometer LISA will be able to monitor the inspiral of stellar-mass black holes around supermassive ones ($10^5 M_\odot$) for up to one year~\cite{AmaroSeoane:2012km, Babak_2017}. These systems, called Extreme Mass Ratio Inspirals (EMRI), will allow us to probe gravity in the strong-field and highly dynamical regime with unprecedented precision.

In order to extract parameters from the signal of the inspiraling objects, one should provide a very accurate template to describe the waveform of the emitted GWs. A simple figure of merit to remember is that the instantaneous GW frequency for LISA should be known with a fractional accuracy of approximately $10^{-8}$ \cite{Thornburg:2011qk}. In GR, the post-Newtonian (PN) formalism allows to compute the waveform of a binary system in the inspiral phase as an expansion in terms of the relative velocity $v$ of the two objects~\cite{Blanchet:2013haa}. In the extreme mass ratio limit, such an expansion can be carried out within black hole perturbation theory~\cite{Poisson:1993aa, Sasaki:1994aa, Thornburg:2011qk}.

Detecting a signal that is not compatible with GR would be an obvious observational breakthrough, but it will be equally important to constrain the possible deviations even if all data remain consistent with GR. One could do so for instance by allowing for independent variations of the PN parameters \cite{LIGOScientific:2019fpa}, as was first done in the parametrized post-Newtonian (PPN) formalism~\cite{Will:1981cz}. Along these lines, the parameterized post-Einsteinian formalism was designed to analyze the leading-order deviations from GR by encoding them in a few  coefficients in the GW amplitude and phase \cite{Yunes_2009}. This kind of parametrized approaches are useful to investigate the consistency of a GR signal \textit{a posteriori}, but  cannot replace the search for a modeled signal beyond GR with an analytic template valid up to high orders in perturbation theory.

At the opposite end of the spectrum, one could choose to study modifications of GR on a model-by-model basis. This approach comes with its own downsides. In fact, it is fair to say that presently there aren't strong theoretical arguments (besides self-consistency and phenomenological viability) to prefer some alternatives to GR over others. The lack of preferred candidates in the landscape of possible gravitational theories makes this strategy impractical and ultimately very inefficient.

A middle-of-the road approach consists in making some broad assumption about the particle content and symmetries of the gravitational sector, and then use effective field theory (EFT) techniques to constrain an entire class of theories. This is the strategy that we will adopt in this paper, and that has been repeatedly used in particle physics as well. In particular, we will focus on theories in which the gravitational sector contains an additional scalar field besides the usual spin-2 graviton---{\it i.e.} on so-called {\it scalar-tensor theories}. Moreover, we will assume that the heavy component of the binary system has a scalar hair---{\it i.e.} a non-trivial radial profile for the scalar field. Such a feature would represent a particularly drastic departure from GR, and thus it is one of the first signatures that one should aim to constrain~\cite{Horbatsch_2012,Berti:2013gfa,Franciolini:2018uyq}.

One might be tempted to immediately invoke a number of no-hair theorems ({\it e.g.}~\cite{Bekenstein:1971hc,Bekenstein:1995un,Hui:2012qt}) to object that scalar charges are quite unusual for black holes. In particular, Hui and Nicolis managed to prove for shift-symmetric scalar tensor theories that the associated Noether current $J^\mu$ must vanish, or else the scalar $J^\mu J_\mu$ would diverge at the black hole horizon~\cite{Hui:2012qt}. However, closer inspection shows that this is not an insurmountable obstacle for scalar field configurations, because one can engineer theories where the current is null and yet black holes admit a non-trivial scalar profile~\cite{Babichev:2017guv}. Also, in some theories such as in scalar Gauss-Bonnet models, the unboundedness of the quantity $J^\mu J_\mu$ does not seem to lead to any pathology for the solution~\cite{Sotiriou:2014pfa}. The more recent Ref.~\cite{Creminelli:2020lxn} further supports the possibility of healthy black hole solutions with scalar hair. Finally, it is well-known that black holes embedded in a scalar gradient can acquire a scalar charge~\cite{Jacobson_1999,Berti:2013gfa,Horbatsch_2012}, and that complex scalars~\cite{Herdeiro:2014goa} or axionic-like particles~\cite{Boskovic:2018lkj} can circumvent the no-hair theorem.  Emboldened by the existence of seemingly healthy asymptotically flat black hole solutions with scalar hair, we believe it is time to go beyond the traditional no-hair paradigm and look for generic imprints that a scalar hair may leave on gravitational wave spectra. 

Analytic waveforms for black holes or neutron stars with a scalar hair are only partially known for some specific theories. For instance, the waveform for ``traditional'' scalar-tensor theories (generalizing the Brans-Dicke theory) is presently known up only to 1PN order~\cite{Lang:2015aa} (although the 2PN calculation is underway~\cite{Bernard_2018, Bernard_2019}). This level of  accuracy would need to be further improved for a meaningful comparison with data, which generally requires the energy flux to be known at least up to 3PN order~\cite{Blanchet:2013haa}. In quadratic gravity (a class of theories that includes among others a coupling to the Gauss-Bonnet term), only the leading PN corrections to the energy flux are known~\cite{Yagi:2011xp}. The post-Einsteinian coefficients are instead currently known for a wide class of modified gravity theories \cite{Tahura_2018}. These few examples have shown that studying the two-body dynamics in modified theories of gravity proves to be an herculean task even in simplest of setups, as the metric and the putative supplementary fields quickly get non-linearly coupled down the PN expansion. 

In this paper, we will develop a general framework to calculate analytical waveforms for EMRIs in scalar-tensor theories under the assumption that the heavy companion features a scalar hair. To this end, we will adopt the EFT for perturbations in the presence of a scalar hair that was first put forward in~\cite{Franciolini:2018uyq} to study departures from the black hole quasi-normal frequencies predicted by GR. This same formalism was also used soon after to prove the existence of stable wormhole configurations with scalar hair~\cite{Franciolini:2018aad}. In Sec.~\ref{sec:recap}, we will review this framework and extend it to include a point-like source. 

The main idea behind this formalism is to take the hairy background solution for granted and focus directly on the dynamics of perturbations. By exploiting the nontrivial background of the scalar field, one can fix radial diffeomorphisms by working in the so-called unitary gauge where the scalar perturbations vanish. Then, the most general Lagrangian for perturbations is only going to be invariant under the residual diffeomorphisms. One major advantage of this approach is that it doesn't require as an input the microscopic Lagrangian of the scalar-tensor theory. As a consequence, it bypasses potential ambiguities associated with field redefinitions of the scalar and conformal (``frame'') transformations of the metric~\cite{Piazza:2013coa}. A similar formalism was initially applied to cosmological perturbations~ \cite{Cheung:2007st,Gubitosi:2012hu,Bloomfield:2012ff} to extract model-independent constraints from observations.

Linear perturbations around spherically symmetric solutions can be classified into even and odd parity modes. The even sector in particular is responsible for some of the most interesting features of GW emission in scalar tensor theories. On the one hand, it is singlehandedly responsible for the leading quadrupole emission (this is the case also in pure GR). Indeed, one byproduct of our computation is that the lowest order radiation in the odd sector is of 1PN order. On the other hand, it contains fluctuations of the scalar field itself, and thus the dipolar component of radiation.  Despite all this, in this paper we will focus our attention on the \emph{odd parity sector}. This is mainly done for reasons of technical simplicity, since this sector contains a single degree of freedom and is described by an effective Lagrangian that includes only a handful of operators. The general formalism that we develop here will be extended to the even parity sector in a future work.

The equations for perturbations of static black hole solutions can be cast into the standard Regge-Wheeler form~\cite{Regge:1957td,Chandrasekhar:1985kt}, characterized by an effective potential $V(r)$ taking values outside the event horizon of the black hole. While quasinormal modes are sensitive to the entire shape of $V$, the PN approximation that we implement here only requires a limited number of terms in a $1/r$ expansion of the potential. In Sec.~\ref{sec-3} we \emph{define} our background metric and the EFT operators by means of such an expansion, and impose  the constraints arising from the tadpole equations to find relations among the coefficients. In Sec.~\ref{sec:examples}, we illustrate our formalism by providing a few examples of covariant theories in the unitary gauge and working out the coefficients of their PN expansion.

To describe EMRI systems we use a point-like source term representing the small mass in circular orbit around the large black hole. We find that the relevant coupling with the odd sector is of the conformal type, and that finite size ({\it i.e.} higher derivative) corrections are negligible with respect to PN ones. This allows us in Sec.~\ref{sec:RW} to write the Regge-Wheeler equation in the presence of a source and express the effective potential $V(r)$ in terms of the PN parameters of the EFT Lagrangian. 

Finally, Sec.~\ref{sec:perturbative_sol_RW} contains the main results of this paper.  Following~\cite{Poisson:1993aa,Sasaki:1994aa}, we  solve the Regge-Wheeler equation in powers $v$. We will work with an accuracy of $\mathcal{O}(v^5)$ beyond the lowest order solution. Since, as emphasized before, the odd sector itself is suppressed by 1PN order with respect to the usual quadrupole formula, 
by calculating the flux at infinity we will obtain an analytic expression for the dissipated power up to 3.5PN order ({\it i.e.}, up to $\mathcal{O}(v^7)$ beyond the leading GR quadrupole), which is the minimal required accuracy for waveform templates.\footnote{Of course, the neglected mass ratio of the system, even if very small in EMRI ($10^{-5}$), will need to be taken into account for an accurate waveform template. Taking into account this mass ratio in a second-order self-force calculation would be an interesting direction in which to extend this work.}
\vspace{.3cm}

\noindent{\it Conventions:} we work in units such that $c =1$ and, throughout most of the paper, we also set $G =1$. This implies that masses and length have the same dimensions. We adopt the ``mostly plus'' metric signature.

\section{Effective theory of black hole perturbations with scalar hair} \label{sec:recap}

In this section, we review the effective theory of perturbations of black holes with scalar hair put forward in~\cite{Franciolini:2018uyq}.  The main idea behind this approach is to assume the existence of a spherically symmetric hairy black hole solution and study the most general dynamics of perturbations that is allowed by symmetry.   

A static black hole solution with a scalar hair features a scalar field with a non-trivial radial profile $\Phi = \bar \Phi (r)$. For our purposes it will not matter whether this is a primary or secondary hair, {\it i.e.} whether or not such profile is associated with an additional conserved charge. Such a profile allows one to study the dynamics of perturbations by working in the so-called {\it unitary gauge}, where the scalar perturbations vanish, {\it i.e.} $\delta \Phi = 0$. This amounts to choosing the radial coordinate $r$ in such a way that hyper-surfaces of constant $r$ coincide with hyper-surfaces of constant $\Phi$. 

The unitary gauge requirement only fixes radial diffeomorphisms. The action in such a gauge must still be invariant under the three residual diffeomorphisms, {\it i.e.} $r$-dependent redefinitions of the time and angular coordinates. This can also be understood from a geometric perspective: the scalar profile introduces a preferred foliation of space-time, and the residual diffeomorphisms correspond to the freedom of specifying a different set of coordinates on each slice of this foliation. We will be particularly interested in four-dimensional quantities associated with this foliation: (1) the coordinate $r$ labeling the various hyper-surfaces, (2) the unit normal vector $n_\mu = \delta_\mu^r / \sqrt{g^{rr}}$, (3) the induced inverse metric $h^{\mu\nu} = g^{\mu\nu} - n^\mu n^\nu$, and (4) the extrinsic curvature tensor $K_{\mu\nu} = h_\mu{}^\lambda \nabla_\lambda n_\nu$.\footnote{Following a widespread notation, we write the extrinsic curvature as a four dimensional tensor (\emph{i.e.} with greek indexes running from $0$ to $3$). As a geometrical object, however, $K_{\mu \nu}$ is indeed three-dimensional, as much as the induced metric $h_{\mu \nu}$. In particular, $K^{\mu r}=K^{r \mu} = 0$. 
} The effective action in unitary gauge can depend on these four  quantities, besides the usual Riemann tensor and covariant derivatives.

It is easy to convince oneself that this geometrical approach in equivalent to a more ``pedestrian'' one, where the effective action is allowed to depend also on $r, g^{rr}$ and $K_{\mu\nu}$, with the understanding that upper $r$-indices do not need to be contracted. Therefore, our effective action in unitary gauge can be written as~\cite{Franciolini:2018uyq} 
\begin{equation} \label{eq:effective action 1}
S = \int d^4 x\sqrt{-g} \, \mathcal{L}\left(g_{\mu\nu}, \epsilon^{\mu\nu\lambda\rho}, 
R_{\mu\nu\alpha\beta}, g^{rr} , K_{\mu\nu} ,\nabla_\mu , r 
\right) \, .
\end{equation}

The effective action \eqref{eq:effective action 1} is not yet optimal to study the dynamics of perturbations.
This is because an infinite number of terms in \eqref{eq:effective action 1} contribute at any given order in perturbations and derivatives. What we are after, instead, is an effective action which makes it manifest that only a finite number of terms are allowed by the symmetries at each order in perturbations and derivatives. To this end, we should decompose the terms appearing in~\eqref{eq:effective action 1} in powers of perturbations, starting with the tadpole terms that fix the background solution. 

To this end, we decompose the building blocks appearing in \eqref{eq:effective action 1} as a sum of a background contribution and a perturbation around it, {\it e.g.} $K_{\mu\nu} \equiv \bar K_{\mu\nu} + \delta K_{\mu\nu}$, and so on. Unlike in the EFT of inflation~\cite{Cheung:2007st} or dark energy~\cite{Gubitosi:2012hu,Bloomfield:2012ff}, this decomposition is not covariant with respect to the residual diffeomorphisms. This subtlety arises because a spherically symmetric background has fewer isometries than an FRW one. For all practical purposes this is however not an issue, as we will discuss at the end of this section. After decomposing all operators in this fashion, one can show that the only terms that contribute to the quadratic action for metric perturbations up to second order in derivatives are: 
\begin{align}
& S =  \int \mathrm{d}^4x \, \sqrt{-g} \bigg[
\frac{1}{2}M^2_1(r) R -\Lambda(r) - f(r)g^{rr} - \alpha(r)\bar K^\mu_{\ \nu} K^\nu_{\ \mu} \nonumber
\\
&	+ M_2^4(r)(\delta g^{rr})^2
	+M_3^3(r) \delta g^{rr }\delta K  + M_4^2(r) \bar K^\mu_{\ \nu} \delta K^\nu_{\ \mu} \delta g^{rr }\nonumber
\\[2mm]
&	+ M_5^2(r)(\partial_r\delta g^{rr})^2
	+M_6^2(r) (\partial_r\delta g^{rr})\delta K  + M_7(r)\bar K^\mu_{\ \nu} \delta K^\nu_{\ \mu}  (\partial_r\delta g^{rr})
	+ M_8^2(r)(\partial_\mu \delta g^{rr})^2 \label{eq: effective action for perturbations}
\\[2mm]
&	+M_9^2(r)(\delta K)^2 + M_{10}^2(r)\delta K^\mu_{\ \nu}\delta K^\nu_{\ \mu}
+ M_{11}(r) \bar K^\mu_{\ \nu} \delta K^\nu_{\ \mu} \delta K 
+ M_{12}(r)  \bar K^\mu_{\ \nu} \delta K^\nu_{\ \rho} \delta K^\rho_{\ \mu}  \nonumber
\\
&	+ \lambda(r) \bar K^\mu_{\ \nu} \bar K^\nu_{\ \rho} \delta K^\rho_{\ \mu} \delta K 
+ M_{13}^2(r) \delta g^{rr }\,  \delta\!\! \  ^{(3)}\!R
+ M_{14}(r)  \bar K^\mu_{\ \nu}\,  \delta\!\! \ ^{(3)}\!R^\nu_{\ \mu} \delta g^{rr } + \ldots
\bigg] \, , \nonumber 
\end{align}
where $^{(3)}R$ is the intrinsic curvature of the hyper-surfaces of constant $r$. Notice that this action contains not only all quadratic terms, but also a subset of higher order terms whose presence is enforced by symmetry considerations. In what follows, we will remain agnostic about the strong coupling scale of this theory and assume that it is high enough that no-linearities and higher derivative corrections can be neglected for our purposes. Notice that the cut-off of the theory is  hidden in the radius-dependent coefficients $M_1(r), \Lambda(r)$, etc..., which contain information about the particular scalar-tensor theory one is considering as well as the hairy black hole solution. For this reason, our assumption can only be  checked on a case-by-case basis.

Some of the functions of $r$ appearing in \eqref{eq: effective action for perturbations} are not arbitrary, but are instead fixed by the requirement that the background metric be a solution to Einstein's equations. More specifically, a spherically symmetric background metric of the form 
\begin{equation}\label{eq:metric} 
\mathrm{d} s^2 = \bar g_{\mu\nu} \mathrm{d} x^\mu \mathrm{d}  x^\nu = -a^2(r)\mathrm{d} t^2 + \frac{\mathrm{d} r^2}{b^2(r)} + c^2(r)\left(\mathrm{d}\theta^2+\sin^2\theta\mathrm{d} \phi^2 \right) \, 
\end{equation} 
will satisfy Einstein's equations only if the coefficients $\Lambda(r), f(r)$ and $\alpha(r)$ appearing in the first line of~\eqref{eq: effective action for perturbations} obey the following \emph{tadpole conditions}:
\begin{align} 
& f(r) 
=  \left(\frac{a'c'}{ac}-\frac{b'c'}{bc} - \frac{c''}{c}  \right)M^2_1 + \frac{1}{2}\left(\frac{a'}{a}-\frac{b'}{b}\right)(M^2_1)' - \frac{1}{2}(M^2_1)'' \label{fm} 
\\
& \qquad \qquad\qquad\qquad\qquad\qquad  - \left( \frac{a'^2}{2a^2}-\frac{a'b'}{2ab}-\frac{a'c'}{ac} + \frac{c'^2}{c^2} - \frac{a''}{2a}\right)\alpha + \frac{a'}{2a}\alpha' \, , \nonumber
 \\[2mm]
&\Lambda(r) 
= - b^2\left(\frac{c''}{c} + \frac{a'c'}{ac}+ \frac{b'c'}{bc}+\frac{c'^2}{c^2}-\frac{1}{b^2c^2}  \right)M^2_1 - b^2\left(\frac{a'}{2a}+\frac{b'}{2b}+\frac{2c'}{c} \right) (M^2_1)'\label{lm} 
\\ 
& \qquad \qquad\qquad\qquad\qquad\qquad  - \frac{b^2}{2}(M^2_1)'' - b^2\left( \frac{a'^2}{2a^2}-\frac{a'b'}{2ab}-\frac{a'c'}{ac} + \frac{c'^2}{c^2} - \frac{a''}{2a}\right)\alpha + \frac{b^2a'}{2a}\alpha'
 \, , \nonumber 
 \\[2mm]
&\left(\frac{a'}{a}-\frac{c'}{c} \right)(M^2_1+\alpha)'  + \left( \frac{a''}{a} - \frac{c''}{c} + \frac{a'b'}{ab} + \frac{a'c'}{ac}- \frac{b'c'}{bc}-\frac{c'^2}{c^2}\right)(M^2_1+\alpha) + \frac{M^2_1}{b^2c^2} = 0\label{thirdtadpcondition}  .  
\end{align}%
Thus, one can think of $\Lambda(r), f(r)$ and $\alpha(r)$ as being completely specified once the background metric and the function $M_1(r)$ are given. All other functions appearing in \eqref{eq: effective action for perturbations} are in principle arbitrary and must be constrained by observations.\footnote{Hairy black hole solutions are notoriously hard to generate, as evidenced by the existence of a variety of ``no-hair theorems'', {\it e.g.}~\cite{Bekenstein:1971hc,Bekenstein:1995un,Hui:2012qt}). Whether or not this difficulty is encoded by additional, hidden constraints on these arbitrary functions---akin to a swampland conjecture for hairy black holes---remains an interesting open question.} 

At this point, we should stress an important difference between the effective action we are using here and the one first derived in \cite{Franciolini:2018uyq}: in Eq. \eqref{eq: effective action for perturbations}, $\bar K^\mu_{\ \nu}$ and $\delta K^\mu_{\ \nu}$ always appear with one upper and one lower index. This was not the case in the effective action used in \cite{Franciolini:2018uyq}. In particular, the tadpole equations appearing in~\cite{Franciolini:2018uyq} imply the specific index structure index structure
$\bar K^{\mu \nu}K_{\mu \nu}$ for the tadpole term proportional to $\alpha(r)$.  As a result our tadpole conditions \eqref{fm}--\eqref{thirdtadpcondition} differ from the ones quoted in~\cite{Franciolini:2018uyq}.  From a conceptual viewpoint the two effective actions are completely equivalent, as they correspond to choosing a different basis of operators in the Lagrangian. From a technical viewpoint, however, the choice we are making here turns out the be more convenient for two reasons. First, matching an explicit scalar-tensor theory onto the effective action \eqref{eq: effective action for perturbations} is much simpler. This is in part due to the fact that the perturbed and background induced metrics with mixed indices have identical components. In App. \ref{sec:matching-example}, we discuss an explicit example to illustrate this point. Second, the transformation properties of $\alpha(r)$ under conformal redefinitions of the metric are much simpler when the third tadpole is defined as in Eq. \eqref{eq: effective action for perturbations}. 

These two advantages amplify when combined with each other. This is because matching calculations are usually easier in the ``Jordan frame'' (where $M_1^2(r)$ is some non-trivial function of the background scalar and the metric), but the dynamics of perturbations is simpler in the ``Einstein frame'' (where $M_1^2(r) =$ constant). As is well known, the two frames are connected by a conformal transformation of the metric. Thus, having EFT coefficients that transform simply under conformal transformations allows us to match in the Jordan frame and then calculate in the Einstein frame. 
  
Finally, let's return to the issue of splitting the operators in \eqref{eq:effective action 1} into background and perturbations. Because this is not a covariant procedure, observational constraints are only meaningful once the gauge has been completely fixed. Thus, some care needs to be taken when comparing constraints derived in different gauges. Otherwise, though, the non-covariant nature of the splitting does not pose any problem. In particular, it can be shown that for any choice of coefficients there exists a Lagrangian of the form \eqref{eq:effective action 1} that exactly reproduces \eqref{eq: effective action for perturbations} up to quadratic order in perturbations~\cite{Franciolini:2018uyq}. Thus, all the terms in \eqref{eq: effective action for perturbations} are actually compatible with the symmetries despite not being explicitly invariant under the residual diffeomorphisms. 

The effective action for perturbations in Eq. \eqref{eq: effective action for perturbations} was used in \cite{Franciolini:2018uyq} to constrain departures from the quasi-normal mode spectrum of Schwarzschild black holes in GR. This EFT framework was also used in~\cite{Franciolini:2018aad} to argue for the existence of stable wormhole solutions in scalar-tensor theories. In remaining of this paper we will develop a third application of this formalism by studying the modifications to the waveform produced by a binary inspiral with extreme mass ratio.


\section{Sourced odd sector in the PN limit} \label{sec-3}

It is convenient to classify perturbations around a spherically symmetric background according to their transformation properties under parity. Because the action \eqref{eq: effective action for perturbations} does not contain parity violating terms, even and odd modes decouple from each other at linear level and can be studied separately. In this paper we restrict our attention to the odd sector of perturbations, which only includes one propagating degree of freedom. The general EFT action~\eqref{eq: effective action for perturbations} is considerably simplified by such a restriction, as only a small subset of operators contributes. In this section, we will first review such a reduced action, and then introduce a parametrization of the EFT coefficients that is appropriate for the PN regime we are ultimately interested in.

\subsection{Effective action for odd perturbations with a point-like source}

Odd metric perturbations are parametrized by 3 functions $h_0$, $h_1$ and $h_2$ as follows~\cite{Regge:1957td}:
\begin{eqnarray} \label{odd metric perturbations}
\delta g_{\mu\nu}^{\rm odd} = \left(
\begin{array}{ccc}
0 & 0 & \epsilon^k {}_j \nabla_k h_0\\
0 & 0 & \epsilon^k {}_j \nabla_k h_1\\
\epsilon^k {}_i \nabla_k h_0 \,\,\,\, &  \epsilon^k {}_i \nabla_k 
                                    h_1 & \,\,\,\, {1\over 2} (\epsilon_i {}^k
                                           \nabla_k \nabla_j +
                                           \epsilon_j {}^k \nabla_k
                                           \nabla_i ) h_2
\end{array}\right) ,
\end{eqnarray}
where
\begin{align} \label{harmonics decomposition odd perturbations}
	h_n (t,r,\theta, \phi) = \sum_{\ell =2}^{\infty} \sum_{m = - \ell}^\ell h_n^{\ell m} (t,r) Y_{\ell m} (\theta, \phi), \qquad \qquad n=1,2,3, 
\end{align}
and $\epsilon_{ij}$ and $\nabla_i$ are respectively the Levi-Civita tensor and covariant derivative on the 2-sphere---see {\it e.g.} Sec. 3 of~\cite{Franciolini:2018uyq} for their explicit expressions. Odd perturbations with angular momentum number $\ell = 0, 1$ have been omitted in Eq. \eqref{harmonics decomposition odd perturbations} because they do not propagate, partly due to an enhanced gauge invariance.  This can be easily checked by deriving the quadratic action for such modes, and is consistent with the fact that the additional scalar degree of freedom belongs to the even sector. 

The only operators in the effective action \eqref{eq: effective action for perturbations} that contribute to the odd sector are~\cite{Franciolini:2018uyq}:
\begin{align}
& S =  \int \mathrm{d}^4x \, \sqrt{-g} \bigg[
\frac{1}{2}M^2_1(r) R -\Lambda(r) - f(r)g^{rr} - \alpha(r)\bar K^\mu_{\ \nu} K^\nu_{\ \mu} \nonumber
\\
& \qquad \qquad \qquad \qquad \qquad + M_{10}^2(r)\delta K^\mu_{\ \nu}\delta K^\nu_{\ \mu}  + M_{12}(r)  \bar K^\mu_{\ \nu} \delta K^\nu_{\ \rho} \delta K^\rho_{\ \mu} \bigg]. \label{S odd}
\end{align}
As shown in the next Section, these terms contribute to the standard Regge-Wheeler equation with two radial and two temporal derivatives. Terms with higher derivatives are not necessarily negligible from the point of view of the PN expansion. However, as they can be \emph{naively} associated with ghost-like instabilities, they must represent small perturbative corrections in the effective Lagrangian. We do not include them here because we assume that they are suppressed by some high scale. 

We now need to supplement this action with a term that describes the coupling with a point-like source. The point-particle action should be invariant under the same symmetries as the bulk action, and for our purposes it will be sufficient to consider only the leading term in its derivative expansion, {\it i.e.}
\begin{align} \label{S source}
	S_{\rm source} = - \int \mathrm{d}\tau \; \mu(r), \qquad\qquad\quad \mathrm{d}\tau^2 = - g_{\mu \nu}\mathrm{d}x^\mu \mathrm{d}x^\nu \; .
\end{align}
In fact, under very reasonable assumptions, higher derivative corrections to the point-particle action above are more suppressed compared to the PN corrections we are interested in. We will discuss this more explicitly in Sec \eqref{sec: finite size} after introducing the PN parametrization of our EFT coefficients.

In order to simplify the subsequent analysis, it is helpful to perform a conformal redefinition of the metric to set $M_1(r) \equiv M_{\rm Pl}$. This can always be achieved provided the arbitrary functions of $r$ appearing in Eqs. \eqref{S odd} and \eqref{S source} are appropriately redefined. We will also extract an overall factor of $(8 \pi)^{-1}$ from all the bulk coefficients, so that out final action reads
 \begin{align}
\begin{split} \label{eq:action-again}
S_\mathrm{odd} = \frac{1}{8\pi} \int \mathrm{d}^4x& \sqrt{-g} \left[\frac{R}{2} - \Lambda(r) -  f(r) g^{rr} - \alpha(r)\bar K^\mu_{\ \nu} K^\nu_{\ \mu}  \right. \\
&  \quad \left. + M_{10}^2(r) \delta K^\mu_{\ \nu}\delta K^\nu_{\ \mu}  +  M_{12}(r)  \bar K^\mu_{\ \nu} \delta K^\nu_{\ \rho} \delta K^\rho_{\ \mu}+ \dots \right] \ - \ \int \mathrm{d}\tau \mu(r) \; .
\end{split}
\end{align}
The last step will allow us to adopt units where $G=1$---which is particularly convenient since we'll be working in the PN regime---without introducing factors of $8 \pi$ in our equations. The fact that our EFT coefficients differ by a factor of $8\pi$ compared to those in~\cite{Franciolini:2018uyq} means that our Einstein equations with $G=1$ should agree with those in~\cite{Franciolini:2018uyq} with $M_{\rm Pl} = 1$.

\subsection{Parametrization of the EFT coefficients} \label{sec:param}

Our next step is to parametrize the background metric components in Eq.~\eqref{eq:metric}, as well as the arbitrary functions of $r$ appearing in the (Einstein-frame) EFT action~\eqref{eq:action-again}. This action (without the source term) was first used in~\cite{Franciolini:2018uyq} to study black hole quasi-normal modes. In that case, the quasi-normal frequencies were found to depend on the values of the EFT coefficients, the metric components, and their derivatives close to the horizon, {\it i.e.} at $r \sim 2M$. Here, in contrast, we will be interested in the PN regime where $r \gg 2M$.

From now on we choose to work with a radial coordinate such that $c(r) = r$ or, equivalently, such that the surface area of the 2-spheres is $4 \pi r^2$. Then, we can expand the remaining components of the background metric in the PN regime as follows 
\begin{subequations}\label{eq:param_a_b}
\begin{align} 
a^2(r) &= 1 - \frac{2M}{r} + a_2 \left(\frac{2M}{r} \right)^2 + a_3 \left(\frac{2M}{r} \right)^3 + \mathcal{O} \left( \frac{M}{r} \right)^4 \; , \\
b^2(r) &= 1 -   \frac{2M}{r} + b_2 \left(\frac{2M}{r} \right)^2 + b_3 \left(\frac{2M}{r} \right)^3 + \mathcal{O} \left( \frac{M}{r} \right)^4 \; .
\end{align}
\end{subequations}
Here, $M$ is the ADM mass of the black hole, which is defined by the first term in the series expansion of $a(r)$. The coefficients $a_i$ and $b_i$  parametrize instead possible deviations from GR. 
We will see later on that terms up to $1/r^3$ are needed to calculate the waveform at $\mathcal{O}(v^5)$ beyond leading order.

One might think that the first few coefficients in the expansion~\eqref{eq:param_a_b} are already tightly constrained by solar system tests. However, the constraints that apply to the Sun do not necessarily apply to other objects. This is particularly evident in theories where the scalar charge is an independent parameter in addition to the mass---in which case the scalar hair is usually called \emph{primary}. Moreover, even in theories with only \emph{secondary} hair,  Birkhoff's theorem does not necessarily apply, and there might exist more than one branch of solutions characterized by different scalar charge/mass ratios.

The reader might also have noticed that $a$ and $b$ are equal up to ${\cal O}(1/r^2)$. Alternatively, a more general expansion for $b$ could  be considered, of the type 
\begin{equation}
 b^2(r) = 1 -  (1+b_1) \frac{2M}{r} + \dots\, .
\end{equation}
In App.~\ref{app-nec} we show that $b_1\neq0$ inevitably leads to violations of the null energy condition (NEC). Theories that violate the NEC are prone to developing instabilities in the scalar sector, although counterexamples also exist ({\it e.g.}~\cite{Creminelli:2010ba,Piazza:2013pua,Franciolini:2018aad}). Allowing for $b_1\neq0$, while considerably complicating the equations, does not seem to lead to any instability in the present case. This is perhaps not surprising, since we are focusing on the odd sector, and instabilities are usually associated with the scalar mode which belongs to the even one. It is possible that also the even sector might remain stable for a suitable choice of operators in the EFT, as was the case for the wormhole solutions studied in~\cite{Franciolini:2018aad}. Nevertheless, establishing this would require a detailed study of the even sector, which we leave for future work. Until then, we will err on the side of caution and restrict ourselves to the case where $b_1=0$, as shown in Eq.~\eqref{eq:param_a_b}.

We now proceed with an analogous expansion for the background coefficients appearing in the EFT action,
\begin{align} \label{exp-1}
\Lambda(r) & =  \frac{2M}{r^3} \left[\Lambda_3 + \Lambda_4 \frac{2M}{r}  + \Lambda_5 \left( \frac{2M}{r} \right)^2 + \dots  \right] \, , \\ \label{exp-2}
f(r) & =  \frac{2M}{r^3} \left[f_3 + f_4 \frac{2M}{r} + f_5 \left(\frac{2M}{r} \right)^2 + \ \dots \ \, \right] \, , \\
\alpha(r) & =  \frac{2M}{r} \left[\alpha_1 + \alpha_2 \frac{2M}{r}  + \alpha_3 \left(\frac{2M}{r} \right)^2 + \ \dots \  \  \label{exp-3} \right] \, , 
\end{align}
Upon use of the  tadpole equations~\eqref{fm}-\eqref{thirdtadpcondition}, the above coefficients can be calculated as polynomials of the metric ones, up to an integration constant that we can choose to be the parameter~$\alpha_1$:

\begin{align}
\Lambda_3  =& -  \alpha_1\, , \qquad \label{3-tadpole1}
\Lambda_4  = \frac{12 a_2 -3 \alpha_1}{4} \, , \\ \label{3-tadpole2}
\Lambda_5  =& \frac{-13 {\alpha_1}+20 {\alpha_1} a_2+24 a_2+30 a_3
-4 {\alpha_1} b_2-8 b_2+6 b_3}{8}\\ \label{3-tadpole3}
 f_3  =& - \alpha_1 \, , \qquad 
 f_4  = \frac{ -7 \alpha_1+4 a_2  + 4 b_2}{4} \, , \\ \label{3-tadpole4}
 f_5 = & \frac{ -27 {\alpha_1}+20 {\alpha_1} {a_2}  
 +4 {\alpha_1} b_2 
+24 {a_2}+6 a_3+8 {b_2}+14 {b_3} }{8}\, ,  \\\label{3-tadpole5}
\alpha_2 =& \frac{6  \alpha_1 -8 a_2}{4} \; , \qquad
\alpha_3 = \frac{\alpha_1(18  - 4  b_2 -12 a_2) - 28 a_2 -18 a_3+4 b_2-2 b_3}{8} . 
 \end{align}

Up to second order in derivatives, the bulk part of the action \eqref{eq:action-again} contains two unknown functions of the radius: $M_{10}(r)$ and $M_{12}(r)$. We will expand also these functions in inverse powers of the radius. A non-zero asymptotic value of $M_{10} (r)$ would induce a non-luminal GW speed at large distances from the black hole. This can be easily seen from the fact that the operator $\delta K_{\mu \nu}$ detunes the radial derivative of $g_{\mu \nu}$ from its temporal derivative (an analogous phenomenon occurs in the EFT of dark energy \cite{ Gleyzes_2013}).  Since the recent detection of a nearly coincident GW and electromagnetic signal constrains deviations from an exactly luminal speed of gravity to be smaller than $ \mathcal{O}(10^{-15})$~\cite{TheLIGOScientific:2017qsa,Creminelli_2017}, we will assume that $M_{10}^2$ starts at order $1/r$ in the PN expansion. It has been argued that the parameters of the EFT of Dark Energy may not be constrained by this measurement~\cite{de_Rham_2018}, because the energy scale at which such theory is defined is very different from the typical GW frequency of a LIGO signal. However, our EFT is precisely designed to study GWs and this bound is particularly relevant. A non-zero asymptotic value of $M_{12}$, on the other hand, does not induce a different speed of gravity in flat space. Therefore, we will parameterize our two EFT coefficients as follows:
\begin{align} \label{M10 expansion}
M_{10}^2 &= \gamma_1 \frac{2M}{r} + \gamma_2 \left( \frac{2M}{r} \right)^2 + \mathcal{O} \left( \frac{M}{r} \right)^3 \; ,  \\
\frac{M_{12}}{M} &= \lambda_0 + \lambda_1 \frac{2M}{r} + \mathcal{O} \left( \frac{M}{r} \right)^2 \; , \label{M12 expansion}
\end{align}
where we have expanded each function at the desired order for our calculations as will be clear in Section \ref{sec:perturbative_sol_RW}. Note that, in units such that $G=1$, $M_{10}$ is dimensionless while $M_{12}$ has mass (or length) dimensions. We have normalized $M_{12}$ by the mass of the central black hole because this choice will simplify later equations. However, it should be kept in mind that $M_{12}$ can also depend on microscopic scales that enter the action of any given scalar-tensor theory (scalar Gauss-Bonnet is an example of this, see Section \ref{sec:examples}). Further observational constraints on the size of the expansion coefficients $\gamma_1$, $\lambda_0$ etc. are discussed in App.~\ref{app:obs}.

Finally, there remains an unknown function in the matter sector introduced by the conformal redefinition of the metric to go in the Einstein frame, $\mu(r)$. This conformal redefinition depends on the background scalar field which is itself expanded in a PN series. We can therefore write an expansion of $\mu$:
\begin{equation}
\mu(r) = \mu_0 \left[1 + \mu_1 \frac{2M}{r} + \mu_2 \left(\frac{2M}{r} \right)^2 + \mu_3 \left(\frac{2M}{r} \right)^3 + \mathcal{O} \left( \frac{M}{r} \right)^4  \right] \; , \label{mu expansion}
\end{equation}
where $\mu_0$ is the asymptotic ADM mass of the point-particle.  
This is reminiscent of the expansion of the mass of an object in terms of the so-called \textit{sensitivities} in Brans-Dicke type theories \cite{Eardley_1975} (similar nonminimal point-particle couplings were also discussed more recently in Einstein-Maxwell-dilaton theories \cite{Juli__2018} or in Einstein-scalar-Gauss-Bonnet gravity \cite{Juli__2019}). Indeed, we will show in App. \ref{app:A} how the coefficients $\mu_i$ are related to the sensitivities in Brans-Dicke type theories.

\subsection{Finite-size corrections to the point-particle action} \label{sec: finite size}

Let us now return to an issue we alluded to earlier, namely the possibility of including higher derivative corrections in the point-particle action \eqref{S source}. As we already pointed out, the point-particle action should be invariant under the same symmetries as the bulk part of the action. This means that, in unitary gauge, all possible terms invariant under residual diffeomorphisms are allowed. These terms will once again be organized in a derivative expansion, and encode the fact that the object under consideration is not truly point-like. 

For example, all the terms\footnote{Notice that the Lagrangian \eqref{finite size Lagrangian} should in principle contain also powers of $u^r$. However, $u^r =0$ for the circular orbits we are interested in, and therefore we have omitted such dependence at the outset for simplicity.} that could in principle contribute a linear coupling with at most one derivative on the metric perturbations are
\begin{align}
	S &= \int d \tau \Big\{ \mu (r) + c_1 (r)  \delta g^{rr} + c_2 (r) \partial^r \delta g^{rr} + c_3 (r) u^\mu \nabla_\mu \delta g^{rr} + c_4 (r) \delta K \nonumber \\
	& \qquad \qquad \qquad \qquad \qquad \qquad \qquad \qquad  + c_5 (r) \bar K^\mu{}_\nu \delta K^\nu{}_\mu + c_6 (r) u_\mu u^\nu \delta K^\mu{}_\nu + \cdots \Big\} , \label{finite size Lagrangian}
\end{align}
where $u^\mu = d x^\mu/d \tau$ is the four-velocity of the point particle, and the dots stand for terms that are further suppressed at large distances (for instance, because they involve additional powers of $\bar K^\mu{}_\nu$).
It turns out that only the very last term in this Lagrangian yields an additional linear coupling with odd perturbations. Using the asymptotic expansions \eqref{eq:param_a_b} and \eqref{mu expansion}, as well as the definition of the angular momentum $L$ in \eqref{eq:def_E_L}, we find that for a circular orbit of radius $r_0$ this last term is schematically of the form
\begin{align}
	u_\mu u^\nu \delta K^\mu{}_\nu \sim \frac{L}{\mu_0 r_0^3} \times \left\{ \begin{array}{c} h_0 \\ h_1 \end{array} \right\}
\end{align}
at leading order in a $1/r_0$ expansion. Thus, assuming that the dimensionless quantity $c_6 (r_0) \sim \mathcal{O}(1)$, as would expected on naturalness grounds, we find that this last term is suppressed compared to the leading interaction \eqref{leading interaction with source} by a factor of $(r_0 \mu_0)^{-1}$. In turn, this correction is negligible compared to the PN corrections we will consider below provided 
\begin{align}
	\frac{1}{r_0 \mu_0} \ll \frac{G M}{r_0} \qquad \Longrightarrow \qquad G M \mu_0 \gg 1 ,
\end{align}
which is certainly satisfied for ``macroscopic'' masses $M$ and $\mu_0$. Thus, in what follows we will neglect finite-size corrections to the point-particle action.

\section{Examples} \label{sec:examples}

To see how the formalism described in the previous sections works in practice, we will now discuss some covariant theories with scalar hair in unitary gauge and calculate the coupling functions of the EFT for these specific examples.

\subsection{Brans-Dicke type theories}

Scalar-tensor theories \emph{\`a la} Brans-Dicke are the simplest playground for modifications of gravity. Black holes cannot display scalar charges in these theories as a consequence of a known and early example no-hair theorem~\cite{Hawking1972}.
However, a neutron star is expected to develop a nonzero scalar charge which can be further enhanced by spontaneous and dynamical scalarization \cite{Damour:1993aa, Barausse_2013}. Of course, we do not expect the waveforms that we will obtain by matching our EFT to a Brans-Dicke type theory to be relevant in the case of a binary neutron star inspiral, where the two components are approximately of the same mass; however, our results could provide a useful cross-check with the existing PN literature \cite{Lang:2015aa, Bernard_2018, Bernard_2019}. Indeed, our waveform should be recovered in the extreme mass ratio limit of these references.
 Furthermore, as was pointed out in~\cite{Cardoso_2013}, a black hole will develop a nonzero scalar charge in realistic astrophysical situations when it is surrounded by matter.

Incidentally,  theories of the Brans-Dicke type provide a nice example of the power of the formalism outlined in the previous sections. For historical reasons, they are often introduced by means of the rather inconvenient action
\begin{equation} \label{eq:bd-action}
S = \frac{1}{16 \pi} \int \mathrm{d}^4x \sqrt{-g} \left[ \phi R - \frac{\omega(\phi)}{\phi} g^{\mu \nu} \partial_\mu \phi \partial_\nu \phi \right] + S_m  \; .
\end{equation}
However, the precise coefficient in front of the Ricci scalar and the functional form of $\omega (\phi)$ are not unambiguously defined, since they can be changed by conformal transformations of the metric and scalar field redefinitions. 
On the contrary, by working in the Einstein frame and in the unitary gauge, our EFT~\eqref{eq:action-again} is free from these ambiguities. 

With an appropriate field redefinition of the metric (\emph{i.e.} by going to the Einstein frame) and of the scalar field, the action~\eqref{eq:bd-action} takes the form (see App.~\ref{app:A} for details and for connection with more standard notation) 
\begin{align} \label{eq:bd_action_EinsteinFrame}
S = \frac{1}{16 \pi} \int \mathrm{d}^4x \sqrt{- \tilde g} \left[ \tilde R -   \tilde g^{\mu \nu} \partial_\mu \varphi \partial_\nu \varphi \right] 
 - \sum_A \int \mathrm{d} t \; m_A(\varphi) \sqrt{-  \tilde g_{\mu \nu} v^\mu_A v^\nu_A} \; ,
\end{align}
where the index $A$ refers to the different point-particle objects,  $v^\mu_A = d x^\mu_A / dt$ is the velocity of each object, and $m_A(\varphi)$ is a field-dependent mass. 
From now on we will drop the tildes for notational convenience. 

For a background field $\bar \varphi(r)$ around a single object of mass $m(\varphi)$, the EFT coefficients in the action \eqref{eq:action-again} are easily identified. Only $f$ and $\mu$ are nonzero and they read
\begin{align}
\begin{split} \label{eq:EFT_coeffs_ST}
f(r) &= \frac{ \bar \varphi'^2(r)}{2} \; , \\
\mu(r) &= m(\bar \varphi(r)) \; .
\end{split}
\end{align}
 In order to obtain a PN expansion of these functions, the only remaining task is to find the background value $\bar \varphi$. By varying the action with respect to $g_{\mu \nu}$ and $\varphi$ one finds the following equations of motion in vacuum
\begin{align}
\begin{split}
&R_{\mu \nu} = \partial_\mu \varphi \partial_\nu \varphi \; , \\
&\partial_\mu \left(\sqrt{-g} g^{\mu \nu}  \partial_\nu \varphi  \right) = 0 \; .
\end{split}
\end{align}
Although an exact solution to these equations is known in the so-called Just coordinates \cite{Damour_1992}, we find it more convenient to solve them perturbatively for large values of $r$ in the standard coordinate system of Eq.~\eqref{eq:metric} with $c(r) = r$. Plugging in our spherically symmetric ansatz for the metric \eqref{eq:metric} and the background field $\bar \varphi(r)$, we find the following system of equations for the three unknown functions $a$, $b$ and $\bar \varphi$ :
\begin{subequations}
\begin{align}\label{eq:system_ST}
 a'' a  + \frac{a' b' a}{b} + \frac{2}{r} a a' &= 0 \; , \\
\frac{a''}{a} + \frac{a' b'}{a b}  + \frac{2b'}{rb} &= - \bar \varphi'^2 \; , \\
r \left( \frac{a' b^2}{a} + b b' \right) + b^2 - 1 &= 0 \; , \\
\partial_r \left(r^2 a b  \bar \varphi'  \right) &= 0 \; .
\end{align}
\end{subequations}
 The last equation is immediately integrated to find
\begin{equation} \label{eq:def_scalar_charge}
\bar \varphi = \varphi_0 - \int \mathrm{d}r \; \frac{q M}{r^2 a b} \; ,
\end{equation}
where $\varphi_0$ is the asymptotic value of the field, $M = m(\varphi_0)$ and $q$ is called the \textit{scalar charge} of the object. For a test-mass, the scalar charge can be easily derived from the coupling function $\omega(\phi)$. For completeness, we include such a computation in App. \ref{app:A}. For a compact object like a neutron star, instead, one should resort to a specific neutron star model encapsulating short-distance physics. In this case, $q$ can grow to appreciable values due to spontaneous scalarization \cite{Damour:1993aa}.

In order to avoid these complications, we choose to treat $q$ as an independent parameter. The remaining task is now to solve perturbativel the first three equations in \eqref{eq:system_ST}. We parametrize $a$ and $b$ as in eq. \eqref{eq:param_a_b}.  It is easy to show that, to be consistent with the order of expansion we are working at, one should expand the two first equations to $\mathcal{O}(1/r^6)$ and the third one to $\mathcal{O}(1/r^4)$ (excluded). We obtain
\begin{align}
\begin{split}
a_2 = 0, \quad b_2 = \frac{q^2}{8}, \quad b_3 = \frac{q^2}{16}, \quad a_3 = \frac{q^2}{48}\, .
\end{split}
\end{align}
The above equations also allow us to find the expansion of $\bar \varphi$
\begin{equation} \label{eq:expansion_barphi}
\bar \varphi = \varphi_0 +\frac{M q}{r} +\frac{M^2 q}{r^2} -\frac{M^3 q \left(q^2-16\right)}{12r^3} + \dots \; .
\end{equation}
However this is not really needed because the coefficients of the effective action are now given through the tadpole equations. In particular, by setting $\alpha_1=0$ we find by using~\eqref{3-tadpole3} and~\eqref{3-tadpole4} that
\begin{equation} \label{eq:f_bd}
f_4 \ = \ \frac{q^2}{8}, \qquad f_5 \ = \ \frac{q^2}{4}\, .
\end{equation}
As expected, the remaining background coefficients, $\Lambda$ and $\alpha$, vanish. 

While referring to App.~\ref{app:A} for more connections with the existing literature, here we point out that the energy flux for brans-dicke type scalar-tensor theories is known up to the 1PN order \cite{Lang:2015aa} (although the 2PN order is underway \cite{Bernard_2018, Bernard_2019}). As already mentioned, when the symmetric mass ratio $\nu = m_1 m_2/(m_1+m_2)^2$ is negligible, we should recover the results of Lang \cite{Lang:2015aa}. However, since we are still missing the even part of the spectrum, we are not able to do this comparison for the time being. We hope to come back to it in a near future.

\subsection{Gauss-Bonnet}

The linear Gauss-Bonnet (GB) scalar tensor model is defined by the following action
\begin{equation} \label{eq:GB-action}
S = \frac{M_P^2}{2} \int d^4 x \sqrt{-g}\left(R - \partial^\mu \phi \partial_\mu \phi  +  2 \bar \alpha  \, \phi \, {\cal G}\right)\, ,
\end{equation}
where we have temporarily re-introduced the reduced Planck mass for later convenience and 
 ${\cal G}$ is the GB  total derivative term,
\begin{equation}
{\cal G} \ =\  R^{\mu \nu \rho \sigma}R_{\mu \nu \rho \sigma} - 4 R^{\mu \nu} R_{\mu \nu} + R^2 \, .
\end{equation}
Following the conventions of~\cite{Sotiriou:2014pfa}, the scalar $\phi$ is dimensionless here, and the coupling of the GB term has been denoted with a bar to distinguish it from the coefficient in the EFT action~\eqref{eq: effective action for perturbations}. Note that EMRIs in the closely related Einstein-dilaton-Gauss-Bonnet and Chern-Simons theories have been studied in~\cite{2016PhRvD..94j4024B} and~\cite{Pani_2011} respectively.

Already from the form of the action \eqref{eq:GB-action} one can see that, if a black hole solution exists for GB gravity, then the scalar field must acquire a non-trivial configuration because it is sourced by a quantity, ${\cal G}$, which itself will be a function of the radius. 

In what follows, we describe the basic steps that one should take to recast the action of this theory in the EFT form~\eqref{eq: effective action for perturbations}. First, it proves useful to exploit the equivalence of GB gravity and a particular Horndeski theory~\cite{Kobayashi:2011nu}. More specifically, the GB term in~\eqref{eq:GB-action} turns out to be equivalent, up to boundary terms, to the Horndeski 5 action~\cite{Horndeski:1974wa,Deffayet:2011gz}
\begin{equation} 
  2 \bar \alpha \, \phi \, {\cal G} \ = \ G_5(X) G_{\mu \nu} \phi^{; \mu \nu} + \frac13 G'_{5}(X)(\Box\phi^3 - 3 \Box \phi \phi_{; \mu \nu} \phi^{;\mu \nu} + 2  \phi_{; \mu \nu} \phi^{;\nu \rho} \phi_{; \rho}^{;\mu}) \ + \ {\rm b. t.}, 
 \end{equation}
with $G_5(X) = - 4 \bar \alpha \ln(|X|)$. In the above, $X \equiv \partial_\mu \phi \partial^\mu \phi$ and $G_{\mu \nu}$ is the Einstein tensor. At this point, we can try adapt the vocabulary of~\cite{Gleyzes_2013} to the present spherically symmetric case in order to write the Horndeski 5 Lagrangian in the EFT form 
~\eqref{eq:GB-action}. To start with, let us focus on the coefficient $M_1^2$  multiplying the Einstein Hilbert term. For the theory at hands it is relatively easy to obtain 
\begin{equation} \label{eq:M1}
M_1^2(r) \ = \ M_P^2\ \left[1 - 8 \bar\alpha b\left(b \phi' \right)'\right]\, .
\end{equation}

The authors of~\cite{Sotiriou:2014pfa} have studied a spherically symmetric black hole solution of the above theory and given the asymptotic behavior at large $r$ of the metric coefficients and of the scalar itself, 
\begin{align} 
a^2(r)&= 1 -\frac{2M}{r} + \frac{M P^2}{6 r^3} +\frac{M^2 P^2+24 \bar \alpha M P }{3 r^4} + {\cal O}(r^{-5})\, , \\ 
b^2(r)&= 1 -\frac{2M}{r} +\frac{P^2}{2 r^2} +\frac{M P^2}{2 r^3} + \frac{48 \bar \alpha M P + 2 M^2 P^2}{3 r^4} + {\cal O}(r^{-5})\, , \\ \label{metricso-3}
c^2(r) & = r^2 \\
\phi(r)& = \frac{P}{r} +\frac{M P}{r^2} +\frac{16 M^2 P-P^3}{12 r^3} +\frac{6 M^3 P -12 \bar \alpha M^2-M P^3}{3 r^4}+ {\cal O}(r^{-5}) \, .
\end{align}
While~\eqref{metricso-3} is just our standard gauge fixing condition,  in the above equations $P$ is the scalar charge of the black hole, which however is not independent of its mass~\cite{Sotiriou:2014pfa}, 
\begin{equation}\label{4-relation-charge}
P = \frac{2 \bar \alpha}{M}\, .
\end{equation}

The above expressions can now be used inside~\eqref{thirdtadpcondition} in order to get an expansion for $\alpha(r)$. Then, by using~\eqref{fm} and~\eqref{lm} we can solve for the remaining tadpole coefficients, $f(r)$ and $\Lambda(r)$. In summary, the effective action coefficients read
\begin{align} \label{M1q-ex}
M_1^2(r) & = 1 - \frac{16 \bar \alpha P}{r^3} - \frac{8 \bar \alpha M P}{r^4} - \frac{4 \bar\alpha(4 M^2 P + P^3)}{r^5} + {\cal O}(r^{-6}) \\ \label{fr-ex}
f(r) & = \frac{P^2}{2 r^4} +\frac{96 \bar \alpha P+2 M P^2}{r^5}+  \frac{240 \bar \alpha M P+24 M^2 P^2-P^4}{4 r^6}+{\cal O}(r^{-7}) \, ,\\ \label{La-ex}
\Lambda(r) & = -\frac{48 \bar \alpha P}{r^5} + \frac{44 \bar \alpha M P}{r^6} + {\cal O}(r^{-7})\, ,\\ \label{alpha-ex}
\alpha(r) & = \frac{24  \bar \alpha P}{r^3} + \frac{8  \bar \alpha M P}{r^4} + {\cal O}(r^{-5})\, .
\end{align}

This is not the end of the story however, because all the quantities above are specified in the ``Jordan frame''. We can always perform a conformal transformation of the metric tensor and bring the action to the form~\eqref{eq:action-again}, with no radius-dependent coefficient multiplying the Einstein Hilbert term. This is achieved by the field redefinition $g_{\mu \nu}^{\rm (J)}(x) \rightarrow g_{\mu \nu}^{\rm (E)}(x) = g_{\mu \nu}^{\rm (J)}(x) M_1^2(r)$, where $J$ and $E$ stand respectively for Jordan and Einstein frames. One can check that equation~\eqref{thirdtadpcondition} is covariant under such a conformal transformation, provided that $\alpha$ transforms as $\alpha^{(J)}(r)\rightarrow \alpha^{(E)}(r) = \alpha^{(J)}(r)/ M_1^2(r)$. Of course, the radial gauge choice $r^2 \equiv g_{\theta\theta} = c(r)$ cannot hold after the conformal transformation. Once we conformally transform to the Einstein frame, we can introduce an ``Einstein frame radius'' $r^2_{(E)} =  g_{\theta\theta} ^{(E)}$ so that the radial gauge choice still holds in the new frame.

However, from the expansion~\eqref{M1q-ex}, it is clear that the conformal factor $M_1^2$ is different from unity only at a relatively high order in the $1/r$ expansion, which makes the Einstein and Jordan frame radii differ only at $\mathcal{O}(r^{-3})$. 
As a result, eqs.~\eqref{fr-ex},~\eqref{La-ex} and~\eqref{alpha-ex} hold also in the Einstein frame. 

Finally, by explicitly translating the Horndeski 5 action into the EFT formalism we obtain the two quadratic operators that are relevant for odd perturbations, 
\begin{align} 
M_{10}^2(r) &=  8 \bar \alpha b^2 \phi'\left(\frac{a'}{2 a} -  \frac{b'}{b} +  \frac{c'}{c} -  \frac{\phi''}{\phi'}\right) \\ \nonumber
& = -\frac{24 \bar \alpha P}{r^3} -\frac{12 \bar \alpha M P}{r^4} -\frac{6 \bar \alpha \left(4 M^2 P+P^3\right)}{r^5} + {\cal O}(r^{-6})\, , \\[2mm]
M_{12}(r) & = - 8 \bar \alpha b \phi' \\ \nonumber
&= \frac{8 \bar \alpha P}{r^2} +\frac{8 \bar \alpha M P}{r^3} + \frac{12 \bar \alpha M^2 P}{r^4}+{\cal O}(r^{-5})\, .
\end{align}
These, however, are too high PN order for the present calculation. In summary, using also relation~\eqref{4-relation-charge}, we obtain for our basic coefficients the following values, 
\begin{align}
\begin{split}\label{eq:GB_coeffs}
& a_2 = 0, \quad a_3 = \frac{P^2}{48 M^2}, \quad b_2 = b_3 = \frac{P^2}{8 M^2}, \quad \mu_3 = \frac{P^2}{2 M^2} , \, \\
& \qquad \quad \alpha_1 = \gamma_1 = \gamma_2 = \lambda_0 = \lambda_1 = \mu_1 = \mu_2 = 0, 
\end{split}
\end{align}

\section{Circular orbits and the sourced Regge-Wheeler equation} \label{sec:RW}

In the extreme mass ratio regime, GWs emitted during the inspiral phase can be thought of as arising from perturbations of the heavy companion generated by a point-like source~\cite{Sasaki:1994aa, Poisson:1993aa}. We can then constrain the presence of a scalar hair for the heavy companion by using the effective theory introduced in the previous section. To this end, we will first discuss some aspects of trajectories in a generic spherically symmetric background, and then derive the Regge-Wheeler equation with a source term.

\subsection{Background trajectories} \label{sec:background_traj}

Before turning our attention to the dynamics of linear perturbations, we will pause for a moment to discuss some features of circular orbits in generic spherically-symmetric backgrounds. The results derived in this section will be used later on to simplify the linear coupling between perturbations and point-like source. Since the metric \eqref{eq:metric} is symmetric around $\theta = \pi/2$, we will consider trajectories that are restricted to this plane. Remember that we are working with a radial coordinate such that $c(r) = r$.

The equations of motion for the point-particle are found by varying the background action,
\begin{equation} \label{eq:matter_action}
\bar S_m = - \int \mathrm{d} \lambda \; \mu(r) \sqrt{\bar g_{\mu \nu} \frac{dx^\mu}{d\lambda} \frac{dx^\nu}{d\lambda}} \; ,
\end{equation}
where we have reintroduced an affine parameter $\lambda$ along the trajectory for convenience.  Denoting $T \equiv \mu(r) \sqrt{\bar{g}_{\mu \nu} \dot{x}^\mu \dot x^\nu}$ and $\dot{x}^\mu \equiv dx^\mu/d\lambda$, the equations of motion take the form
\begin{equation}
\frac{d}{d\lambda} \frac{\partial T}{\partial \dot x^\mu} = \frac{\partial T}{\partial x^\mu} \; .
\end{equation}
Choosing now the affine parameter $\lambda$ to be equal to the background proper time $\tau$, and using the fact that $T$ does not depend explicitly on $t$ nor on $\phi$, we obtain the two equations
\begin{align}
\frac{d}{d\tau} \left(\mu(r) a^2(r) \frac{dt}{d\tau} \right) = 0 \qquad \qquad 
\qquad \frac{d}{d\tau} \left(\mu(r) r^2 \frac{d\phi}{d\tau} \right) = 0 \; .
\end{align}
The conserved quantities in parentheses are respectively the energy and the angular momentum of the point-particle:
\begin{equation} \label{eq:def_E_L}
E = \mu(r) a^2(r) \frac{dt}{d\tau}, \qquad\qquad\qquad L = \mu(r) r^2 \frac{d\phi}{d\tau} \; .
\end{equation}
Using $d \tau^2 = - \bar{g}_{\mu \nu} dx^\mu dx^\nu$, one finds that the radial component of the equations of motion can be written using the conserved quantities above as
\begin{equation}\label{1st order eq circular orbits}
\left( \frac{dr}{d\tau} \right)^2 = b^2 \left[ \frac{E^2}{\mu^2 a^2} - \left(1+\frac{L^2}{\mu^2 r^2} \right)  \right] \; .
\end{equation}
Deriving now Eq. \eqref{1st order eq circular orbits} with respect to $\tau$, we find the second order equation
\begin{equation}  \label{2nd order eq circular orbits}
2 \frac{d^2r}{d\tau^2} = \frac{d}{dr}\left[ \frac{E^2 b^2}{\mu^2 a^2} - b^2 \left(1+\frac{L^2}{\mu^2 r^2} \right)  \right] \; .
\end{equation}

In the particular case of a circular trajectory, for which $dr/d\tau = d^2r/d\tau^2 = 0$, we can solve Eqs. \eqref{1st order eq circular orbits} and \eqref{2nd order eq circular orbits} to find the energy and the angular momentum of the particle as a function of the background parameters evaluated at the radius of the orbit:
\begin{align} \label{eq:L_E_background}
L = \mu r \; \sqrt{\frac{r(a \mu)'}{\mu(a-r a')}} , \qquad \qquad \qquad E = \mu a \; \sqrt{\frac{\mu a+r a \mu'}{\mu(a-r a')}} \; ,
\end{align}
where $( \,\, )'\equiv d/dr ( \,\, )$. It is easy to check that these results reduce to the usual expressions for angular momentum and energy of a non-relativistic point particle when $\mu(r) = m$, $a(r)^2 = 1 - 2 M/r$, and $M/r = v^2 \ll 1$  . It is also interesting to notice that, generically, there will be corrections to Kepler's law (we will come back to this point in Section \ref{sec:last_part}). Indeed, from Eqs. \eqref{eq:def_E_L} and the definition of the angular frequency $\Omega = d\phi/dt$ we find the relation
\begin{equation} \label{Omega eq.}
\Omega = \frac{d\phi}{dt} = \frac{L a^2}{E r_0^2} \; ,
\end{equation}
with $r_0$ the radius of the orbit. In GR, it is easily checked that $\Omega^2 r_0^3 = M$ for a Schwarzschild solution in coordinates such that $4 \pi r^2$ is the area of the invariant 2-spheres. In general, this will no longer be true in the presence of a scalar hair.

\subsection{Sourced Regge-Wheeler equation} \label{sec:RW_with_source}

We are finally in a position to derive the linearized equation for odd metric perturbations sourced by a test particle. In what follows, we will work in Regge-Wheeler gauge by setting $h_2 =0$ (see eq.~\ref{odd metric perturbations}). Perturbations with different values of $\ell$ and $|m|$ decouple at linear level due to rotational invariance. Therefore, we will focus on a single $(\ell, |m|)$ sector and suppress the angular momentum labels whenever possible to simplify the notation. By expanding the bulk part of the effective action \eqref{eq:action-again} up to quadratic order in perturbations we find 
\begin{align} \label{bulk action 1}
	S_\mathrm{bulk} &= \frac{1}{8\pi} \int \mathrm{d} t \mathrm{d} r \left[u_1 |h_0|^2 + u_2 |h_1|^2 + u_3( |\dot h_1|^2 - 2 \dot h_1^* h_0' + |h_0'|^2 + 2u_4\dot h_1^* h_0) + \text{c.c.} \right],
\end{align}
with the understanding that for $m=0$ one should add an overall factor of $1/2$ to avoid overcounting.
The real functions $u_i(r)$ were calculated in~\cite{Franciolini:2018uyq}, and are reproduced here in Appendix~\ref{app:def_complicated_functions} for completeness. 

The bulk action should be supplemented with the point-particle action expanded up to linear order in perturbations. Considering again a circular trajectory in the $\theta = \pi /2$ plane with radius $r_0$ and  angular frequency $\Omega$, and choosing the affine parameter so that $\lambda = t$, we find the following expression for the source action

\begin{align} \label{leading interaction with source}
	S_\mathrm{source} &= - \frac{s_{\ell m} L}{r_0^2} \int \mathrm{d} t \; h_0 (t,r_0) e^{i m \Omega t} + \text{c.c} ,
\end{align}
where $L$ is the angular momentum defined in Eq. \eqref{eq:def_E_L}, and the coupling constant $s_{\ell m}$ is given by
\begin{align} \label{PPP}
	s_{\ell m} =\, \sqrt{\frac{2\ell +1}{4\pi} \frac{(\ell-m)!}{(\ell +m)!}} \, P_\ell^{m+1} (0) \ ,
\end{align}
with $P_\ell^{m+1}$ an associated Legendre polynomial.

The perturbation $h_0$ never appears with a time derivative  in the total action $S=S_{\rm bulk} + S_{\rm source}$. Therefore, it can be integrated out by solving its equation of motion, which is just a constraint equation. This equation is however a second-order ordinary differential equation in the $r$ variable. To overcome this difficulty, we will follow Refs.~\cite{Franciolini:2018uyq, De_Felice_2011} and introduce an auxiliary variable $q(t,r)$ defined by
\begin{equation} \label{eq:def_q}
q = \dot h_1 - h_0' + u_4 h_0 \; .
\end{equation}
This variable can be thought of as an extension of the Regge-Wheeler variable to black hole solutions with a scalar hair. In fact, one can check explicitly that, up to an overall coefficient specified below in Eq. \eqref{Psi def}, $q$ reduces precisely to the standard Regge-Wheeler variable in the absence of a scalar hair and for a Schwarzschild background metric. Using Eq. \eqref{eq:def_q}, we can now rewrite the total action as
\begin{align}
S &= \frac{1}{8\pi} \int \mathrm{d} t \mathrm{d} r \biggr\lbrace (u_1 - \partial_r(u_3 u_4) - u_3 u_4^2) |h_0|^2 + u_2 |h_1|^2 +  u_3 q^*\left[2 (\dot h_1 - h_0' + u_4 h_0) - q \right] \biggr\rbrace  \nonumber \\
& \qquad \qquad \qquad \qquad  \qquad \qquad \qquad \qquad \qquad \ - \frac{s_{\ell m} L}{r_0^2} \int \mathrm{d} t \; h_0 e^{i m \Omega t} + \text{c.c.} \; . \label{total action 1}
\end{align}
It is easy to show that the bulk part of the action is equivalent to the one in Eq.~\eqref{bulk action 1} after integrating out~$q$. Varying instead with respect to $h_0^*$ and $h_1^*$ one obtains the algebraic constraints
\begin{equation}
h_0 = \frac{\partial_r(u_3 q) + u_3 u_4 q}{\partial_r(u_3 u_4) + u_4^2 u_3 - u_1} + A(r) \delta(r-r_0) e^{im \Omega t} \; ,\qquad \qquad \quad h_1 = \frac{u_3}{u_2} \dot q \; . \label{sols 0 h1}
\end{equation}
Once again $r_0$ is the radius of the circular orbit, whereas $A(r)$ is defined by
\begin{equation}
A(r) =  \frac{8 \pi s_{\ell m} L}{2r^2(u_1-\partial_r(u_3 u_4) - u_4^2 u_3)} \; .
\end{equation}
Plugging the solutions \eqref{sols 0 h1} for $h_0$ and $h_1$ into the action \eqref{total action 1}, we obtain (up to an irrelevant divergent constant) the effective action
\begin{align}
S &= \frac{1}{8\pi} \int \mathrm{d} t \mathrm{d} r \left[ \mathcal{G}_{00} \dot q^2 + \mathcal{G}_{rr} q'^2 + \mathcal{G}_{qq} q^2 \right] \nonumber  \; \\
& \qquad \qquad \qquad \qquad \quad + \frac{1}{4\pi} \int \mathrm{d}t \mathrm{d}r \; u_3 \left [ (u_4 A - A')\delta(r-r_0) - A \delta'(r-r_0) \right ] q \, e^{i m \Omega t} , \label{q action}
\end{align}
where the explicit expressions of the $\mathcal{G}$ functions are given in Appendix \ref{app:def_complicated_functions}. By varying the quadratic action above with respect to $q$ we obtain a second order equation for the only propagating degree of freedom in the odd sector: 
\begin{equation}
- \frac{\mathcal{G}_{rr}}{\mathcal{G}_{00}} q'' - \frac{\partial_r \mathcal{G}_{rr}}{\mathcal{G}_{00}} q' + \left(\omega^2 + \frac{\mathcal{G}_{qq}}{\mathcal{G}_{00}} \right) q = -  \frac{u_3}{\mathcal{G}_{00}} \left [ (u_4 A - A')\delta(r-r_0) - A \delta'(r-r_0) \right ] 2 \pi \delta(\omega - m \Omega) \; , \label{eq. odd perts 1}
\end{equation}
where we have transformed to Fourier space (in time) with the convention $h(\omega) = \int \mathrm{d} t \; e^{i \omega t} h(t)$. This equation can be recast in a Schr\"odinger-like form by rescaling $Q$ and redefining the radial coordinate as follows:
\begin{subequations}
\begin{align}
 \Psi &= ( \vert \mathcal{G}_{rr} \mathcal{G}_{00} \vert)^{1/4} q \label{Psi def} \\
 \frac{d}{d\tilde r} &= \sqrt{\left \vert \frac{\mathcal{G}_{rr}}{\mathcal{G}_{00}} \right \vert} \frac{d}{dr} \; \label{eq:def_tortoise_coord} ,
\end{align}
\end{subequations}
Then, Eq. \eqref{eq. odd perts 1} reduces to the Regge-Wheeler equation
\begin{equation} \label{eq:RW_with_source}
\frac{d^2 \Psi}{d \tilde r^2} + \left[\omega^2 - V(\tilde r)\right] \Psi = J(\tilde r) \; ,
\end{equation}
with
\begin{align}\label{eq:source_term}
J &= 2 \pi \delta(\omega - m \Omega) u_3 \frac{ \vert \mathcal{G}_{rr} \mathcal{G}_{00} \vert^{1/4}}{\mathcal{G}_{00}} \left[ (A'-u_4A) \delta(r-r_0) + A \delta'(r-r_0)\right] \; , \\
V &= \frac{5 \mathcal{G}_{00}'{}^2
   \mathcal{G}_{rr}{}^2-\mathcal{G}_{00}{}^2 \left(16
   \mathcal{G}_{qq}
   \mathcal{G}_{rr}-\mathcal{G}_{rr}'{}^2+4
   \mathcal{G}_{rr}
   \mathcal{G}_{rr}''\right)-2 \mathcal{G}_{00}
   \mathcal{G}_{rr}
   \left(\mathcal{G}_{00}'
   \mathcal{G}_{rr}'+2
   \mathcal{G}_{00}''
   \mathcal{G}_{rr}\right)
   }{1
   6 \mathcal{G}_{00}{}^3
   \mathcal{G}_{rr}} \ . \label{eq:def_RW_potential}
\end{align}

\section{Power emitted in the PN regime} \label{sec:perturbative_sol_RW}

In this section we derive the main result of our paper: a PN expansion of the power emitted in the odd sector by an extreme mass-ratio binary in the presence of a scalar hair. Our approach will be similar to the one first developed in~\cite{Poisson:1993aa} and then used in~\cite{Sasaki:1994aa, Tagoshi_1994} to calculate the power dissipated by a particle in circular orbit around a Schwarzschild black hole  up to the 4PN order. We organize our discussion in several steps: first, we show how the power emitted is related to the form of $\Psi$ far away from the source (Sec. \ref{sec:GW_fluw}); second, we review how the asymptotic form of $\Psi$ is related to the homogeneous solution $\Psi^\mathrm{in}$ satisfying ingoing boundary conditions at the horizon (Sec. \ref{subsec:setup}); third, we calculate $\Psi^\mathrm{in}$ in a PN expansion (Sec. \ref{sec-6.2}); and finally, we assemble all our results to obtain a PN expansion of the power emitted (Sec. \ref{sec:last_part}).

\subsection{Dissipated power from asymptotic solution} \label{sec:GW_fluw}

We will start by expressing the energy flux leaving the binary system in terms of the variable $\Psi$ defined by Eqs. \eqref{eq:def_q} and \eqref{Psi def}. To this end, we will make the simplifying assumption that our quadratic action for perturbations reduces to the one we would derive from GR at large distances. This assumption is certainly well supported by present observations---{\it e.g.}, by the current bounds on the luminal propagation of speed of GWs~\cite{Creminelli_2017}---and it can be translated into a bound on the asymptotic behavior of the coefficients in the effective action \eqref{eq:action-again}: both $M_{10}(r)$ and $M_{12}(r)/r$ must vanish at large~$r$.  As we will see in Sec.~\ref{sec:examples}, this assumption is also easily satisfied by known black hole solutions with scalar hair.

We can now combine the explicit expressions provided in App.~\ref{app:def_complicated_functions} together with the PN expansions \eqref{eq:param_a_b} to show that at large $r$
\begin{equation} \label{estimates}
u_4 \to \frac{2}{r}, \qquad\qquad\quad   ( \vert \mathcal{G}_{rr} \mathcal{G}_{00} \vert)^{1/4} \to \sqrt{\frac{\ell(\ell+1)}{(\ell-1)(\ell+2)}} \frac{r}{2} \; ,
\end{equation}
and therefore that our variable $\Psi$ reduces asymptotically to 
\begin{equation}
	\Psi \to \frac{r}{(\ell+2)(\ell -1)} \, [ \dot h_1 - r^2 \partial_r(h_0/r^2)] \ .
\end{equation}
Up to an overall multiplicative factor that only depends on $\ell$, this  result is equal to the usual Regge-Wheeler variable, see {\it e.g.} ~\cite{Nagar:2005ea}. Thus, we can immediately borrow standard results to relate the asymptotic form of $\Psi$ to the usual $+$ and $\times$ polarizations in flat space. Keeping track of the aforementioned overall factor, we have~\cite{Nagar:2005ea}
\begin{equation} \label{h plus cross to Psi}
h_+ - i h_\times = \frac{2 i}{r} \; _{-2} Y_{\ell m} (\theta, \phi) \Psi \; ,
\end{equation}
where $ \; _{-2} Y_{\ell m}$ are the $s=-2$ spin-weighted spherical harmonics. Finally, because the asymptotic action for $h_+$ and $h_\times$ is the same as in GR, so will be the asymptotic form of the stress-energy tensor of perturbations. Thus, the total instantaneous power emitted takes the usual form in terms of $h_{+,\times}$~\cite{Maggiore:1900zz}:
\begin{align} \label{power emitted general}
	P \ = \ \lim_{r \to \infty} \int d\Omega \, \frac{r^2}{16 \pi} ( \dot h_+^2 +  \dot h_\times^2) \ = \ \lim_{r \to \infty} \ \frac{1}{4\pi} \ \sum_{\ell \geq 2} \sum_{m=-\ell}^\ell \left \vert \frac{d \Psi}{dt} \right \vert^2  ,
\end{align}
where in the last step we used the orthonormality of spin-weighted spherical harmonics and assumed that $h_+ - i h_\times$ is a linear superposition of modes with all possible values of $\ell, m$.

We will use this result later in this section to calculate the total power emitted in the odd sector in a PN expansion. For now, we just point out that the corresponding expression in the even sector would be more complicated because it would need to include also the power emitted by the additional  scalar mode. We leave a full investigation of the even sector for future work.

\subsection{Asymptotic solution from homogenous solution} \label{subsec:setup}

Equation \eqref{power emitted general} means that the power emitted is completely determined by the asymptotic form of the solution $\Psi$ to the inhomogeneous equation \eqref{eq:RW_with_source}. Following~\cite{Sasaki:1994aa}, we will now show how the latter can in turn be expressed in terms of one particular solution to the associated homogeneous equation. 

As long as the ``potential'' $V(\tilde r)$ vanishes at the horizon ($\tilde r \to - \infty$) and at spatial infinity ($\tilde r \to + \infty$), $\Psi$ will asymptotically approach a linear combination of complex exponentials $e^{\pm i \omega \tilde r}$ (notice that $J(\tilde r)$ vanishes at both ends). On physical grounds, we will impose ingoing boundary conditions at the horizon---corresponding to the fact that no classical signal can leave the horizon---and outgoing boundary conditions at spatial infinity---since we are not interested in gravitational radiation produced by other, far-away sources. Using the Green's functions method, we can express such a solution as follows: 
\begin{align} \label{Green function solution}
\Psi(\tilde r) &= \int_{-\infty}^\infty d\tilde r' G(\tilde r, \tilde r') J(\tilde r') \; . 
\end{align}
The Green's function $G$ with the correct boundary conditions is in turn equal to
\begin{align} \label{Feynman Green's function}
	G(\tilde r,\tilde r') &= \frac{1}{W}  \left[ \theta(\tilde r-\tilde r') \Psi^\mathrm{out}(\tilde r) \Psi^\mathrm{in}(\tilde r') + \theta(\tilde r'-\tilde r) \Psi^\mathrm{out}(\tilde r') \Psi^\mathrm{in}(\tilde r) \right]\; , 
\end{align}
with $W = \Psi^\mathrm{in} \partial_{\tilde{r}} \Psi^\mathrm{out} - \Psi^\mathrm{out} \partial_{\tilde{r}} \Psi^\mathrm{in}$ the Wronskian,
$\theta$ the Heaviside function, and $\Psi^\mathrm{out}$ and $\Psi^\mathrm{in}$ solutions to the homogeneous Regge-Wheeler equation subject to ingoing and outgoing boundary conditions respectively:
\begin{subequations}\label{eq:BC}
\begin{align}  \label{Psi in bc}
\Psi^\mathrm{in} &= \left\{
\begin{array}{ll}
C e^{-i \omega \tilde{r}}, & \qquad \mathrm{for} \; \tilde{r} \rightarrow - \infty \\
A^\mathrm{in} e^{-i \omega \tilde{r}} + A^\mathrm{out} e^{i \omega \tilde{r}}, & \qquad \mathrm{for} \; \tilde r \rightarrow +\infty
\end{array} \right. \\
\Psi^\mathrm{out} &= \left\{
\begin{array}{ll}
B^\mathrm{in} e^{-i \omega \tilde{r}} + B^\mathrm{out} e^{i \omega \tilde{r}}, & \qquad \mathrm{for} \; \tilde{r} \to - \infty \\
e^{i \omega \tilde{r}}, & \qquad \mathrm{for} \; \tilde r \rightarrow +\infty \\
\end{array} \right. .
\end{align}
\end{subequations}
Notice that we have fixed the normalization of $\Psi^\mathrm{out}$ at $\tilde r \rightarrow + \infty$, but have kept the normalization of $\Psi^\mathrm{in}$ arbitrary for later convenience. 

Since the Wronskian is constant by construction, we can evaluate it using the asymptotic expressions \eqref{eq:BC}, which at spatial infinity yield $W = 2 i \omega A^\mathrm{in}$. Taking now the limit $\tilde r \rightarrow + \infty$ of \eqref{Green function solution}, we obtain an expression for $\Psi$ far away from the emission region:
\begin{equation} \label{eq:solution_RW_with_source}
\Psi(\tilde r) \rightarrow \frac{e^{i \omega \tilde r}}{2i\omega A^\mathrm{in}} \left[ \int_{-\infty}^\infty d\tilde r' \, \Psi^\mathrm{in}(\tilde r') J(\tilde r') \right]  \; .
\end{equation}
Thus, we see that the amplitude of the emitted wave is completely determined by the ingoing-wave solution of the homogeneous Regge-Wheeler equation and its related coefficient $A^\mathrm{in}$. Of course, $\Psi^{\rm in}$ and therefore $A^{\rm in}$ are determined up to an overall multiplicative constant, but this ambiguity does not affect Eq. \eqref{eq:solution_RW_with_source}, which only depends on the ratio $\Psi^{\rm in}/A^\mathrm{in}$.

\subsection{PN expansion of the homogeneous solution} \label{sec-6.2}

Let us now calculate $\Psi^{\rm in}$ and $A^{\rm in}$ by solving the homogeneous Regge-Wheeler equation in a PN scheme with the ingoing boundary conditions \eqref{Psi in bc}. Given how we parametrized the various functions entering the Regge-Wheeler equation in Section \ref{sec:param}, we will find it more convenient to work in terms of the coordinate $r$ rather than the tortoise coordinate $\tilde r$. Even better, we will introduce a dimensionless coordinate $z \equiv \omega r$, so that our homogeneous equation becomes:
\begin{equation} \label{eq:RW}
 \frac{dz}{d\tilde r} \frac{d}{dz} \left( \frac{dz}{d\tilde r}  \frac{d \Psi^\mathrm{in}}{d z } \right) + (\omega^2 -V) \Psi^\mathrm{in} = 0 \; .
\end{equation}
The advantage of working with a dimensionless coordinate is that the two scales entering this equation, $M$ and $\omega$, can only appear in the dimensionless combination $\epsilon \equiv  2 M \omega$ (remember, our units are such that $G = 1$). Moreover, the source $J$ in \eqref{eq:source_term} is non-zero only for $\omega = m \Omega$, and in this regime we have 
\begin{align}
	\epsilon \sim 2M \Omega = \frac{2M}{r_0} \times r_0 \Omega = v^3 \ll 1,
\end{align}
where $r_0$ is once again the radius of the circular orbit. This suggests that we solve Eq. \eqref{eq:RW} in perturbation theory by expanding in powers of $\epsilon$. More precisely, we will write our equation~as 
\begin{equation}
E_0[\Psi^\mathrm{in}] + \epsilon E_1[\Psi^\mathrm{in}] + \epsilon^2 E_2[\Psi^\mathrm{in}] + \mathcal{O}(\epsilon^3) = 0 \; ,
\end{equation}
and look for a perturbative solution of the form
\begin{equation} \label{eq:expansion_psi}
\Psi^\mathrm{in} = \Psi^\mathrm{in}_0 + \epsilon \Psi^\mathrm{in}_1 + \epsilon^2 \Psi^\mathrm{in}_2 + \mathcal{O}(\epsilon^3) \; ,
\end{equation}
satisfying ingoing boundary conditions at the horizon. Similarly, we expand $A^\mathrm{in}$ in powers of $\epsilon$ as $A^\mathrm{in} = A^\mathrm{in}_0 + \epsilon A^\mathrm{in}_1 + \epsilon^2 A^\mathrm{in}_2 + \mathcal{O}(\epsilon^3)$. We will see later on that working up to $\mathcal{O}(\epsilon^2)$ included is sufficient to calculate the waveform up to $\mathcal{O}(v^5)$ beyond the leading order result. 

In order to expand Eq. \eqref{eq:RW} in powers of $\epsilon$, we will use the following results, which can be derived from those in App.~\ref{app:def_complicated_functions}:
\begin{align}
\frac{d\tilde r}{dr} &= 1 + \kappa_1 \frac{\epsilon}{z} + \kappa_2 \frac{\epsilon^2}{z^2} + \mathcal{O} \left( \epsilon^3 \right) \; , \label{PN expansion tortoise coordinate}\\
V(z)   &= \frac{\omega^2}{z^2} \bigg(  \ell(\ell+1) + \kappa_3 \frac{\epsilon}{z} + \kappa_4 \frac{\epsilon^2}{z^2} + \mathcal{O} \left( \epsilon^3 \right) \bigg) \; ,
\end{align}
where
\begin{subequations}\label{eq:def_kappa}
\begin{align} 
\kappa_1 &= 1 + \gamma_1 + \frac{\lambda_0}{2} \; , \\
\kappa_2 &= \frac{1}{8(\ell -1)(\ell + 2)}  \big[  8 a_2 - 8 b_2 - (2 \alpha_1 + 4 \gamma_1 + \lambda_0)^2 \nonumber \\
 & +(\ell^2 + \ell - 2) (8 - 4 a_2 - 4 b_2 + 8 \gamma_1 + 12 \gamma_1^2 + 8 \gamma_2 + 
 2 \lambda_0 + 12 \gamma_1 \lambda_0 + 3 \lambda_0^2 + 
 4 \lambda_1)  \big]  \vphantom{\frac{1}{2}} \; , \\
 \kappa_3 &= -6 + 4 \alpha_1 + 5 \gamma_1 - (\ell^2 + \ell - 3) \bigg( 1 + \frac{\lambda_0}{2} \bigg)  \; ,
\end{align}
\begin{align}
 \kappa_4 &=  \frac{1}{ 16(\ell -1)(\ell + 2)} \bigg\lbrace \ell (\ell+1) \big[ \ell( \ell+1) \big( 16 a_2 - 4 \lambda_0 (-4 + 4 \gamma_1 + \lambda_0) - 8 \lambda_1 \big) \nonumber \vphantom{\frac{1}{2}} \\
&-336 a_2 - 8 \alpha_1 \lambda_0 + 76 \gamma_1 \lambda_0 + 
 11 \lambda_0^2 + 
 4 \big( 12 + 52 b_2 + 4 (-4 + \alpha_1) \alpha_1 - 62 \gamma_1 \nonumber \vphantom{\frac{1}{2}} \\
 & + 
    24 \alpha_1 \gamma_1 + 35 \gamma_1^2 + 36 \gamma_2  - 
    23 \lambda_0 + 16 \lambda_1 \big) \big] + 8 \big[ 54 a_2 - 30 b_2 + 7 \alpha_1^2 + 
   62 \gamma_1 \nonumber \vphantom{\frac{1}{2}} \\
   & + (\gamma_1 + \lambda_0) (9 \gamma_1 + 
      2 \lambda_0) + \alpha_1 (16 + 20 \gamma_1 + 13 \lambda_0) - 
   3 (4 + 12 \gamma_2 - 5 \lambda_0 + 4 \lambda_1) \big] \vphantom{\frac{1}{2}} \bigg\rbrace \; .
\end{align}
\end{subequations}
In the GR limit, where all our EFT parameters are set to zero, we recover the usual tortoise coordinate and Regge-Wheeler potential. Note that a similar PN parametrization of the Regge-Wheeler potential was discussed in \cite{2019PhRvD..99j4077C, 2019PhRvD.100d4061M}. However, here we are able to express the parameters $\kappa_i$ in terms of more fundamental ones.

\subsubsection{Zeroth-order solution: $\Psi^\mathrm{in}_0$} \label{subsec:psi0}

At lowest order in perturbation theory, the equation we need to solve is 
\begin{align}
	E_0[\Psi_0^\mathrm{in}] \equiv \Psi_0^\mathrm{in}\hphantom{}''(z) + \left(1-\frac{\ell(\ell+1)}{z^2}\right) \Psi_0^\mathrm{in}(z) = 0 .
\end{align}
The equation for $\Psi_0^{\rm in}$ is the same as in GR and there is no dependence on modified gravity parameters. This simple equation admits the two independent solutions $z j_\ell(z)$ and $z y_\ell(z)$, where $j_\ell$ and $y_\ell$ are spherical Bessel functions. The ingoing boundary condition should be imposed at the horizon $z_* \sim M \omega \sim \epsilon$. At zeroth order in our perturbative expansion, consistency requires that we set $z_* = 0$.  Our solutions scale like $z j_\ell(z) \sim z^{\ell+1}$ and $z y_\ell(z)\sim z^{-\ell}$ for small $z$,  and therefore regularity at the horizon singles out the solution
\begin{align}
	\Psi_0^\mathrm{in}(z) = z j_\ell(z) .
\end{align}
We can now calculate the zeroth order contribution to the coefficient $A^{\rm in}$ using the well-known asymptotic form of the spherical Bessel function $j_\ell$, which gives:
\begin{align}
	\Psi_0^\mathrm{in}(z) \ \stackrel{z \to \infty}{\longrightarrow} \ \frac{i}{2} \left( e^{-i z} e^{i \ell \pi/2} - e^{i z} e^{-i \ell \pi/2} \right) .
\end{align}
Using the fact that $z = \omega r \simeq \omega (\tilde r+ \mathcal \varphi)$ asymptotically, with $\varphi$ an arbitrary integration constant, we conclude that $A^{\rm in}_0 = i^{\ell+1} e^{i \varphi} / 2 $. Thus,  we find that $A^{\mathrm{in}}$ is determined up to an arbitrary phase, in agreement with the remarks of Ref.~\cite{Sasaki:1994aa}. This phase has no physical meaning, as evidenced by the fact that the power emitted ultimately depends on $|A^{\rm in}|^2$.

Notice that at lowest order the precise position of the horizon is irrelevant: our solution is valid provided the horizon is at $r_* = \mathcal{O}(1) M$. In fact,  the PN order at which the distinction between $z_* =0$ and $z_* \sim \epsilon$ becomes important is determined by the size of  $\Psi_0^\mathrm{in}(\epsilon) /A^{\rm in}_0 \sim \epsilon^{\ell+1} \sim v^{3 \ell + 3}$. For the lowest multipole, $\ell = 2$, this yields a contribution to the power emitted of order $|J|^2 \times \mathcal{O} (v^{18})$, given that the power emitted scales like $|\Psi|^2$. This is consistent with the result derived by Sasaki in pure GR~\cite{Sasaki:1994aa}, and is beyond the PN order we will be considering here---especially given that $J$ itself scales like some positive powers of $v$. Incidentally, the absence of outgoing modes at the horizon becomes relevant at the exact same order. Up until that point, regularity determines our solution uniquely.

\subsubsection{First-order solution: $\Psi^\mathrm{in}_1$} \label{subsec:psi1}

Let us know turn our attention to the first order correction to the solution above. We now need to solve an inhomogeneous equation for $\Psi_1^{\rm in}$ of the form $E_0 (\Psi_1^{\rm in}) = - E_1 (\Psi_0^{\rm in})$, where the explicit form of the source term is
\begin{align}
	E_1[\Psi_0^\mathrm{in}] &= -2 \frac{\kappa_1}{z} \Psi_0^\mathrm{in}\hphantom{}''(z) + \frac{\kappa_1}{z^2} \Psi_0^\mathrm{in}\hphantom{}'(z)  - \frac{\kappa_3}{z^3}  \Psi_0^\mathrm{in}(z) \; ,
\end{align}
where the $\kappa_i$'s are defined in Eq.~\eqref{eq:def_kappa}.

Once again, the first order solution can be found using the Green's functions method, which yields
\begin{align} \label{Psi 1}
	\Psi_1^{\rm in}(z) = - \int_0^\infty d z' G_0 (z, z') E_1 (\Psi_0^{\rm in}(z')),
\end{align}
with $G_0$ a Green's function of the differential equation $E_0$ which satisfy the appropriate boundary conditions. 
We want to make sure that lowest order solution $\Psi_0^{\rm in}$ is regular at the horizon, and we want to make sure not to spoil this. For this reason, we will choose $G_0$ in such a way that $\Psi_1^{\rm in}(z=0) =0$. This is accomplished by the Green function
\begin{align}
	 G_0 (z, z') = \theta (z-z') \left[ z y_\ell(z)  z' j_\ell(z') - z j_\ell(z) z' y_\ell(z') \right] ,
\end{align}
where we used the fact that the Wronskian of $E_0$ is $W = z j_\ell \partial_z (z y_\ell) - z y_\ell \partial_z (z j_\ell) = 	1$. The Green function $G_0$ is the analog of a retarded Green's function (with time replaced by the radial coordinate $z$), unlike the one in Eq. \eqref{Feynman Green's function}, which is closer in spirit to a Feynman's Green function. 

Taking now the large-$z$ limit of \eqref{Psi 1}, we can read off the coefficient of the exponential $e^{-i \omega \tilde r}$ to extract the first order correction to $A^{\rm in}$:\footnote{The naive large-$z$ limit of \eqref{Psi 1} diverges logarithmically. This divergence however cancels against a subleading term in the asymptotic expansion of $\Psi_0^{\rm in} (r)$ that arises when $r$ is expressed in terms of $\tilde r$ up to next-to-leading order in $\epsilon$.} 
\begin{align} \label{A^in_1}
A^{\rm in}_1 &=  \frac{i^{\ell+2}e^{i \varphi}}{4} \bigg[ \frac{\lambda_0 + 4 (\alpha_1 + \gamma_1 - 1)}{\ell(\ell+1)} - 1 - \frac{\lambda_0}{2} + (2 + 2 \gamma_1 + \lambda_0) (1 + \psi_\ell - \ln(2))
\bigg] \;  , 
\end{align}
where $\psi_\ell$ is the digamma function defined by
\begin{equation}
\psi_\ell = \sum_{k=1}^{\ell-1} \frac{1}{k} - \gamma \; ,
\end{equation}
and $\gamma = 0.577...$ is Euler's constant. 

The careful reader may worry about the fact that our result \eqref{A^in_1} does not seem to reduce to the GR result (see {\it e.g.} Eq. (4.6) of Ref. \cite{Sasaki:1994aa}) when all our EFT parameters vanish. This discrepancy is however due to a different choice of normalization and phase for $\Psi$. As stated in Ref. \cite{Sasaki:1994aa}, only the difference between the value of $A^{\rm in}_1$ for different $\ell$'s is physically meaningful, and indeed we have checked that such difference correctly reproduces the GR result in the appropriate limit.

\subsubsection{Second-order solution: $\Psi^\mathrm{in}_2$} \label{subsec:psi2}

Extending our first order analysis to higher orders is conceptually straightfoward. In particular, the equation for the second order correction $\Psi_2^{\rm in}$ is of the form $E_0[\Psi^\mathrm{in}_2] = - E_1[\Psi^\mathrm{in}_1] - E_2[\Psi^\mathrm{in}_0]$, with 
\begin{align}
	E_2[\Psi^\mathrm{in}_0] =  \frac{3 \kappa_1^2 - 2 \kappa_2 }{z^2} \Psi_0^\mathrm{in}\hphantom{}''(z) - \frac{3 \kappa_1^2 - 2 \kappa_2}{z^3} \Psi_0^\mathrm{in}\hphantom{}'(z)  - \frac{\kappa_4}{z^4}  \Psi_0^\mathrm{in}(z) \; ,
\end{align}
Following the same logic we adopted to derive the first order solution, we conclude immediately that 
\begin{align}
	\Psi_2^{\rm in}(z) = - \int_0^\infty d z' G_0 (z, z') [E_1 (\Psi_1^{\rm in}(z')) + E_2 (\Psi_0^{\rm in}(z'))].
\end{align}
As we will discuss in a moment, we will not need $A_2^{\rm in}$ to calculate the power emitted at the PN order we'll be interested in---the small-$z$ limit of $\Psi_2^{\rm in}(z)$ will be sufficient.

\subsection{PN expansion of the dissipated power}\label{sec:last_part}

We are now finally in a position to calculate the power emitted by combining all the results we have derived so far in this section. Combining the explicit expression \eqref{eq:source_term} for the source term $J$ with the asymptotic form of $\Psi^{\rm in}$ in Eq. \eqref{eq:solution_RW_with_source}, Fourier transforming back from frequency to time, and then plugging the result into the formula  \eqref{power emitted general} for the power emitted, we find
\begin{align}
	P = \frac{1}{16 \pi} \sum_{\ell \geq 2} \sum_{m=-\ell}^\ell \left\vert \left(\frac{A'}{A} - u_4 - \frac{d}{dr} \right) \left( \frac{\Psi^{\rm in}}{A^{\rm in}} A \, u_3 \frac{ \vert \mathcal{G}_{rr} \mathcal{G}_{00} \vert^{1/4}}{\mathcal{G}_{00}}  \frac{d \tilde r}{dr}\right) \right\vert^2_{r=r_0} .
\end{align}
Thus, we see that the power emitted depends  on $\Psi^{\rm in}$, the background metric coefficients, and all other EFT coefficients evaluated only at the radius $r_0$ of the orbit. This means in particular that we can expand all coefficients in powers of $2M /r_0 \sim v^2$, and $\Psi^{\rm in}(z_0)$ (which itself is an expansion in $\epsilon \sim v^3$) in powers of $z_0 \equiv m \Omega r_0 = m v$. When taken all together, these expansions will yield a PN approximation for the power $P$.

In order to carry out these expansions systematically, we need to take into account the fact that Kepler's law---and thus the relation between $r_0$ and $v$---is modified in the presence of a scalar hair. In fact, the velocity is equal to 
\begin{equation}\label{eq:Kepler}
v = \Omega r_0 = \frac{L a^2}{E r_0} \; ,
\end{equation}
where in the last step we used eq. \eqref{Omega eq.}, while the energy $E$ and angular momentum $L$ are defined in Eq.  \eqref{eq:L_E_background}. Expanding $E$ and $L$ in inverse powers of $r_0$, we can invert this equation to express $2M/r_0$ in terms of $v$:
\begin{align} \label{eq:kepler}
	\frac{2M}{r_0} = \frac{ v^2}{1-2\mu_1} \left[1 + \frac{4a_2-6\mu_1+8\mu_2}{(1-2\mu_1)^2} v^2 + \mathcal{O}(v^4) \right].
\end{align}
This implies that the parameter $\epsilon = 2 M m \Omega = (2 M / r_0) m v $ also admits an expansion in powers of $v$. Incidentally, the fact that the leading term on the righthand side depends on the parameter $\mu_1$ can be viewed as a renormalization of Newton's constant---a common feature in scalar-tensor theories \cite{Damour_1992, Kuntz:2019zef}.

The results derived in this section are sufficient to calculate the dissipated power up to $\mathcal{O}(v^5)$ beyond the leading order result. The limiting factor is the expansion of $\Psi^{\rm in}$ in powers of $\epsilon$, which we have carried out only up to second order. Expanding simultaneously in powers of $\epsilon$ and $z_0$ yields an expression of the form            
\begin{equation}
\frac{\Psi^\mathrm{in} (z_0)}{z_0^{\ell+1}} \sim  [1 + \mathcal{O}(z_0^2)] + \epsilon \left[ \frac{1}{z_0} + \mathcal{O}(z_0) \right] + \epsilon^2 \left[ \frac{1}{z_0^2} + \mathcal{O}(1) \right] +\epsilon^3 \left[ \frac{1}{z_0^3} + \mathcal{O}(1/z_0) \right] + \mathcal{O}(\epsilon^4) \; .
\end{equation}
Therefore, the contribution cubic in $\epsilon$ that we are neglecting would give a leading correction that scales like $(\epsilon / z_0)^3 \sim v^6$. The second order correction to $A^{\rm in}$ would contribute at the same order, which is why we have not calculated it.

Our PN expansion for the power emitted in the odd sector can be cast in the form   
\begin{align}
\begin{split}\label{eq:power}
P &= P_N v^2 \Big[ p_{0} \;  + p_{2} \; v^2 + p_{4} \; v^4 + \mathcal{O}(v^6) \Big]  \; ,
\end{split}
\end{align}
where we have denoted by $P_N$ the standard quadrupole energy loss,
\begin{equation}
P_N = \frac{32}{5} \left( \frac{\mu_0}{M} \right)^2 v^{10} \; ,
\end{equation}
and the coefficients $p_i$ in Eq. \eqref{eq:power} are reported in App.~\ref{power-app}. This is the main result of our paper: our EFT approach provides a model-independent parametrization of how the coefficients $p_i$ can differ from their GR value in the presence of a scalar hair.  A departure from GR would lead to changes in the phase of the waveform. 

An interesting byproduct of our result is that we can turn off all our EFT coefficients to obtain the power emitted in the odd sector in GR. To our knowledge, this is the first time that the contribution of the odd sector to the luminosity formula for a circular orbit in GR has been appeared in the literature (earlier work was based on the Teukolsky equation and yielded results that included both even and odd sector).  In particular, our result shows that the odd sector contribution is already of 1PN order---{\it i.e.} suppressed by $v ^2$ compared to the quadrupole expression, which comes from the even sector. Thus, we have determined the total power emitted in the odd sector up to 3.5PN order.

One final comment we should make, related to our discussion about boundary conditions at the horizon in Sec. \ref{subsec:psi0}, is that Eq.~\eqref{power emitted general} only accounts for the power dissipated in GWs at \textit{infinity}. However, a fraction of the GWs emitted will also be absorbed by the black hole horizon. Fortunately, this effect is again of higher PN order compared to the accuracy considered in this paper. This can be seen from the near-horizon behavior of the full solution $\Psi$ in Eq.~\eqref{Green function solution},
\begin{equation} \label{eq:solution_RW_with_source_horizon}
\Psi(\tilde r) \xrightarrow[\tilde r \to - \infty]{} \frac{C e^{-i \omega \tilde r}}{2i\omega A^\mathrm{in}} \left[ \int_{-\infty}^\infty d\tilde r' \, \Psi^\mathrm{out}(\tilde r') J(\tilde r') \right]  \; .
\end{equation}
On the one hand, we have that $C/ A^{\rm in} \sim \Psi_0^\mathrm{in}(\epsilon) /A^{\rm in}_0 \sim \epsilon^{\ell+1}$; on the other hand, $\Psi^\mathrm{out}$ must be (at lowest order) a combination of spherical Bessel functions $z j_\ell(z)$ and $z y_\ell(z)$ in order to enforce purely outgoing boundary conditions at infinity. Thus, $\Psi^\mathrm{out} (z_0) \sim z_0^{-\ell}$ for $z_0 \ll 1$, and therefore the ratio of the solution to the Regge-Wheeler equation at the two boundaries is
\begin{equation}
\frac{\Psi(\tilde r \rightarrow - \infty)}{\Psi(\tilde r \rightarrow  \infty)} \sim v^{\ell+2} \; .
\end{equation}
Since $\ell \geq 2$ and the power is proportional to $\vert \Psi \vert^2$, this means that the power dissipated at the horizon is suppressed by $v^8$ compared to the power emitted at infinity. This is consistent with the results derived by Poisson and Sasaki in pure GR~\cite{Poisson_1995}.

\section{Conclusions} \label{sec:conclusions}

This new era of GW observations should not catch unprepared those looking for new physics. 
Among other sources,  extreme mass ratio inspiral systems (typically, a solar mass black hole orbiting around a supermassive one) represent one of the main targets of the future space-borne interferometer LISA and will provide some exquisite test of GR in the strong field regime~\cite{AmaroSeoane:2012km}. By describing the small companion as a test particle, these systems are suited to efficient analytic treatment by using black hole perturbation theory in the presence of a source~\cite{Sasaki_2003}.

In theories other than GR though, the bank of templates to fit against GW signals is still relatively restricted. In Brans-Dicke type theories---the simplest scalar-tensor alternatives to GR---waveforms of binary neutron stars are only known to the 1PN order \cite{Lang:2015aa, Bernard_2018, Bernard_2019}, which is insufficient for data analysis.
At the same time, when looking for alternatives to GR in order to produce new templates, one should decide at which level of detail to work. 

In this paper we have explored and endorsed an EFT formalism that can apply to any scalar-tensor model allowing 
a non-trivial profile for the scalar around the black hole. The dynamics of the perturbations is distilled in a certain number of coupling functions of the radius that are unambiguous against possible field redefinitions, both in the scalar and in the metric sectors. We have modeled the source with a point-particle action representing the small BH perturbing the background spacetime. This allowed us to derive a modified Regge-Wheeler equation ruling the evolution of gravitational perturbations.  Solving this equation in the PN regime gives access to the power emitted from the system. For simplicity, we have focused on the odd sector of perturbations which is technically simpler. Extending our formalism to the even sector should be straightforward, albeit computationally more challenging than the odd one. 

Our final and most important formula is the power dissipated in the odd modes up to 3.5PN order, Eq.~\eqref{eq:power}. It depends on a set of parameters which represent deviations of the metric and of the action from GR in the PN regime. In this sense, our formalism could be viewed as a parametrized PN framework applied to hairy BHs.

Given our final equation~\eqref{eq:power} for the dissipated power, one may be tempted to ask what is the advantage of our formalism compared to e.g letting $p_0$, $p_1$ and $p_2$ free in an actual template, as this is done in the parametrized post-Einsteinian (ppE) formalism \cite{Yunes_2009}. Our approach can be seen as a convenient bridge between theoretical models and observations, analogous to what is normally called phenomenology in particle physics.  In this paper we have related the ``observed''  parameters $p_0$, $p_1$ and $p_2$  to the expansion coefficients ($a_2, b_2, \alpha_1, \mu_1$ etc.) of a given theory (see App.~\ref{power-app} for the explicit formulae). Examples of how to derive such expansion coefficients are given in Sec.~\ref{sec:examples}.
So once a particular theory is chosen, say with one free modified gravity parameter, then our EFT provides a one-parameter family of templates which can be used to draw more conclusive tests of GR than the ppE formalism can provide due to its inherent freedom of parametrization. We think that this method can give rise to interesting and efficient modeled searches in a large class of modified gravity theories.

\acknowledgments

The authors would like to thank Vitor Cardoso, Walter Goldberger, Ira Rothstein, Luca Santoni, Filippo Vernizzi, and Scott Watson for helpful discussions. We are also grateful to Emanuele Berti, Vitor Cardoso, and Luca Santoni for comments on the manuscript. R.P. is supported in part by the National Science Foundation under Grant No. PHY-1915611.

\appendix

\section{Decomposition of $K_{\mu \nu} K^{\nu \rho} K_\rho^{\ \mu}$} \label{sec:matching-example}

In this appendix we provide an example to illustrate why it is convenient to use mixed indices for the extrinsic curvature. In order to raise and lower the indices of the unperturbed  extrinsic curvature tensor $\bar K^\mu_{\ \nu}$, it would seem natural to use the background metric. However, when dealing also with perturbations, this is far from being the most convenient option. A quantity such as $\delta K^\mu_{\ \nu} = K^\mu_{\ \nu} - \bar K^\mu_{\ \nu}$ would end up transforming in some hybrid cumbersome way. In practice, when trying to translate a general theory in the EFT language, one has to expand in perturbations terms such as $K_{\mu \nu} K^{\nu \rho} K_\rho^{\ \mu}$, and would like to be able to raise and lower the indices in some definite standard way at any step of the process. In the case of a spatially flat Friedmann-Robertson-Walker (FRW) universe~\cite{Piazza:2013coa}, the extrinsic curvature of the constant time hyper-surfaces evaluates $\bar K^\mu_{\ \nu} = {\rm diag}(0, H, H, H) = H h^\mu_{\ \nu}$, $H$ being the Hubble parameter and $h^\mu_{\ \nu}$ the induced (full, \emph{i.e.} containing the perturbations!) three-dimensional metric. One can thus \emph{define} the perturbations as the fully covariant tensor $\delta K_{\mu \nu} \equiv K_{\mu \nu} - H h_{\mu \nu}$ and raise and lower the indices accordingly~\cite{Cheung:2007st}. 
 
In the present less symmetric case, there is no natural way to define a covariant perturbation tensor $\delta K^\mu_{\ \nu}$. By looking at the  metric in the form
\begin{equation}\label{eq:metric2} 
\mathrm{d} s^2 = \bar g_{\mu\nu} \mathrm{d} x^\mu \mathrm{d}  x^\nu = -a^2(r)\mathrm{d} t^2 + \frac{\mathrm{d} r^2}{b^2(r)} + c^2(r)\left(\mathrm{d}\theta^2+\sin^2\theta\mathrm{d} \phi^2 \right) \, 
\end{equation} 
one finds
\begin{equation}
 \bar K^\mu_{\ \nu} \ = \ b(r) \cdot {\rm diag}\left(\frac{a'(r)}{a(r)},\ 0, \ \frac{c'(r)}{c(r)}, \ \frac{c'(r)}{c(r)}\right)\, .
\end{equation}

In the absence of a covariant perturbation tensor, it is misleading to manipulate terms containing background and perturbation quantities. One solution is to just abstain to do so, by always contracting extrinsic curvature tensors with an index up and an index down, without the need of ever lowering and raising indices. For example, when expanding in perturbations, the cubic extrinsic curvature term previously mentioned should here be written as 
\begin{align}
K^{\mu}_{\  \nu}\,  K^\nu_{\  \rho}\,  K^\rho_{\ \mu} \ & = \ (\bar K^{\mu}_{\  \nu} + \delta K^{\mu}_{\  \nu}) \, (\bar K^\nu_{\  \rho} + \delta K^\nu_{\  \rho})\,  (\bar K^\rho_{\ \mu} + \delta K^\rho_{\ \mu})\\ \nonumber
& = \ - 2 \bar K^{\mu}_{\  \nu}\, \bar K^\nu_{\  \rho}\, \bar K^\rho_{\ \mu} + 3 \bar K^{\mu}_{\  \nu}\, \bar K^\nu_{\  \rho}\, K^\rho_{\ \mu} + 3
\bar K^{\mu}_{\  \nu}\, \delta K^\nu_{\  \rho}\, \delta K^\rho_{\ \mu} + \delta K^{\mu}_{\  \nu}\, \delta K^\nu_{\  \rho}\, \delta K^\rho_{\ \mu} \, ,
\end{align}
\emph{i.e.} with each term having one index up and one index down. 
A similar reasoning applies to the induced intrinsic curvature, $\  ^{(3)}\!\bar R^\mu_{\ \nu} = {\rm diag}\left(0, 0, c^{-2}(r), c^{-2}(r)\right)$. Contractions involving more indices, such as those involving covariant derivatives of $K^\mu_{\ \nu}$, will require more care, but they appear at higher order in the derivative expansion and can be overlooked at this time. 

With the caution required by the issues just discussed, one can follow the construction of~\cite{Franciolini:2018uyq} 
and show that the only terms that contribute to the quadratic action for metric perturbations up to second order in derivatives are those appearing in~\eqref{eq: effective action for perturbations}  

\section{The null energy condition} \label{app-nec}

Consider a spherically symmetric metric in the form
\begin{equation}
ds^2 = - a^2(r) dt^2 + \frac{dr^2}{b^2(r)} + r^2 d\Omega\, ,
\end{equation}
and let us expand the coefficients $a$ and $b$ in general powers of $1/r$, 
\begin{subequations}\label{eq:param-app}
\begin{align} 
a^2(r) &= 1 - \frac{2M}{r} + a_2 \left(\frac{2M}{r} \right)^2 + a_3 \left(\frac{2M}{r} \right)^3 + \mathcal{O} \left( \frac{M}{r} \right)^4 \; , \\
b^2(r) &= 1 -  (1+b_1) \frac{2M}{r} + b_2 \left(\frac{2M}{r} \right)^2 + b_3 \left(\frac{2M}{r} \right)^3 + \mathcal{O} \left( \frac{M}{r} \right)^4 \; .
\end{align}
\end{subequations}
We want to study under which conditions the null energy condition (NEC) is satisfied. The NEC asserts that, for any null vector $u$, 
\begin{equation}
T_{\mu \nu} u^\mu u^\nu \ \geq 0\, . \label{nec}
\end{equation}

In the Einstein frame, the components of the energy momentum tensor can be calculated simply by imposing the Einstein equations. For the metric~\eqref{eq:param-app} they read
\begin{align}
T_{tt} & \ = \ - a^2 b^2 \left(\frac{1}{r^2} + \frac{2 b'}{r b} - \frac{1}{r^2 b^2}\right)\, , \\
T_{rr} & \ = \ \left( \frac{1}{r^2} + \frac{2 a'}{r a} - \frac{1}{r^2 b^2}\right)\, , \\
T_{\theta \theta} & \ = \ r^2 b^2 \left(\frac{a''}{a} + \frac{a' b'}{ab} + \frac{a' c'}{ac} +\frac{ b' c'}{bc} \right)\, .
\end{align}
By spherical symmetry, we can always point our null vector towards, say, a point of the sphere of symmetry at $\phi = 0$. Moreover, the null vector is defined up to an overall constant. Finally, the condition of being null reduces the degree of freedom of such a vector to just one independent parameter $B$, as follows, 
\begin{equation}
u^\mu \ = \ \left(\frac{1}{a},\ bB, \ \frac{\sqrt{1-B^2}}{r}, \ 0\right)\, ,
\end{equation}
where $B$ goes from $0$ (tangential direction) to $1$ (pure radial direction).

We now calculate the quantity in~\eqref{nec} at each order in $1/r$. The first non trivial order is at $1/r^3$, 
\begin{equation} 
T_{\mu \nu} u^\mu u^\nu \ = \ \frac{2 M b_1 (1 - 3B^2)}{r^3} \ + \ \dots\, .
\end{equation}
The above expression changes sign at $B = 1/\sqrt{3}$. The only way it can be non-negative is $b_1=0$. If we impose that, then at the next order we have
\begin{equation} 
T_{\mu \nu} u^\mu u^\nu \ = \ \frac{8 M^2}{r^4} \left[ a_2 + B^2 (b_2 - 2 a_2)\right]\ + \dots \, ,
\end{equation}
which is always positive for $b_2>a_2>0$.
We have checked that at higher order one has to deal with similar inequalities among the relevant coefficients $a_i$, $b_i$. It is just at $1/r^3$ order that NEC implies a specific value of the parameter (\emph{i.e.}, $b_1 = 0$).

\section{Observational constraints from time delays} \label{app:obs}

From eq. \eqref{PN expansion tortoise coordinate}  we could infer that there is an additional Shapiro time delay induced by the parameters $\lambda_0$ and $\gamma_1$ on a GW as compared to a photon. As the recent GW170817 event has remarkably set a stringent bound on the speed of gravitons compared to photons \cite{TheLIGOScientific:2017qsa,Creminelli_2017}, let us see the consequences of this measurement in our formalism (even if the GW170817 event concerns two neutrons stars of comparable mass where our perturbative treatment is expected to break down, we are just concerned here by orders of magnitude). Such a violation of the equivalence principle was already used in \cite{Abbott_2017} to constrain the difference $\left \vert \gamma_\mathrm{GW} - \gamma_\mathrm{EM} \right \vert$ between the Eddington parameters of GW and photons respectively.

For an external observer, a photon traveling on a radial geodesic of the metric \eqref{eq:metric} has an apparent ``speed''
\footnote{The photon speed is unchanged by the conformal transformation in the matter sector of eq. \eqref{eq:action-again}. However, one should be careful to the fact that the mass appearing in Eq.~\eqref{eq:drdt_phot} is not the physical mass that one would measure with a gravitational experiment. As suggested by the form of Kepler's law \eqref{eq:kepler}, this physical mass is related to $M$ \textit{via} $M_\mathrm{phys} = M (1-2 \mu_1)$.  }
\begin{equation} \label{eq:drdt_phot}
\frac{dr}{dt} = a b \simeq 1 - 2 \frac{M}{r} + \mathcal{O}\left( \frac{M}{r} \right)^2 \; .
\end{equation}
If the photon starts its trajectory from near the horizon of the central black hole and ends it in a detector on Earth, the Shapiro time delay (defined as the delay between the time of flight of this photon and the time of an equivalent photon traveling in a Minkowski spacetime \cite{Shapiro:1964aa}) is
\begin{equation}
\Delta t_\mathrm{photon} = - 2 M \log \left( \frac{d}{r_s} \right) \; .
\end{equation}
where $d$ is the distance of the black hole to the Earth, and $r_s$ is its Schwarzschild radius. Note that this time is finite while we would expect a true signal emitted near to the horizon to be infinitely redshifted: this is due to the first-order approximation in $1/r$ which we are using. On the other hand, a GW is a solution to the RW equation \eqref{eq:RW_with_source} and its wavelike properties are associated to the coordinate $\tilde r$. From Eq. \eqref{PN expansion tortoise coordinate} the delay for a graviton then reads
\begin{equation}
\Delta t_\mathrm{graviton} = - (2+\lambda_0+2\gamma_1) M \log \left( \frac{d}{r_s} \right) \; .
\end{equation}
For nonzero $\lambda_0$, $\gamma_1$ there is a difference in the time of arrival of photons and gravitons. However, the order of magnitude of this time difference is (in natural units) the Schwarzschild radius of the central object. This is way below the $1.7$s time difference measured in the GW170817 event for a solar mass object, even if log-enhanced by the ratio $d/r_s$; however this could become relevant for a supermassive black hole. We conclude that the current bound on the GW speed is not currently constraining the parameters of our expansion.

\section{Matching to Brans-Dicke like theories} \label{app:A}

In this Appendix, we will connect our results for Brans-Dicke type theories with the existing PN literature \cite{Lang:2015aa, Bernard_2018, Bernard_2019}. In these references, the two fundamental quantities are the Brans-Dicke coupling function $\omega(\phi)$ and the field-dependent mass $m_A(\phi)$, defined by the Brans-Dicke like action (in the Jordan frame)
\begin{equation} \label{eq:bd_action_app}
S = \frac{1}{16 \pi} \int \mathrm{d}^4x \sqrt{-g} \left[ \phi R - \frac{\omega(\phi)}{\phi} g^{\mu \nu} \partial_\mu \phi \partial_\nu \phi \right] - \sum_A \int \mathrm{d} t \; m_A(\phi) \sqrt{- g_{\mu \nu} v^\mu_A v^\nu_A} \; ,
\end{equation}
where the index $A$ refers to the different point-particle objects,  $v^\mu_A = d x^\mu_A / dt$ is the velocity of each object, and $m_A(\phi)$ is a field-dependent mass. 
The field dependence of $m_A(\phi)$ is expected since local physics now depends on the environment and so of the value of $\phi$ at the object location \cite{Eardley_1975}. For light objects, we do not expect any dependence of $m_A$ on $\phi$, while for strongly self-gravitating objects like neutron stars we cannot ignore this dependence. This phenomenon is known as spontaneous scalarization~\cite{Damour:1993aa}. Following \cite{Mirshekari_2013, Bernard_2018} the coupling function $\omega(\phi)$ is parameterized in a weak-field expansion as
\begin{equation} \label{eq:expansion_omega}
\omega(\phi) = \frac{1-4\zeta}{2 \zeta} + \omega_1 \frac{1-\zeta}{\zeta^2} \left( \frac{\phi}{\phi_0} - 1 \right) +  \frac{\omega_2}{2} \frac{1-\zeta}{\zeta^3} \left( \frac{\phi}{\phi_0} - 1 \right)^2 + \dots \; ,
\end{equation}
where $\phi_0$ is the asymptotic value of the scalar, and the mass function $m_A(\phi)$ as
\begin{equation} \label{eq:expansion_mA}
\log m_A(\phi) = \log m_A^{(0)} + s_A \log \frac{\phi}{\phi_0} + \frac{s_A'}{2} \left( \log  \frac{\phi}{\phi_0} \right)^2 + \frac{s_A''}{6} \left( \log  \frac{\phi}{\phi_0} \right)^3 + \dots
\end{equation}

We can express the Jordan frame action~\eqref{eq:bd_action_app} in our formalism by making the conformal field redefinition
\begin{equation} \label{eq:change_variable_ST}
\tilde g_{\mu \nu} = \phi g_{\mu \nu}, \quad \phi = e^\Phi, \quad \varphi = \int d\Phi \left( \omega(e^\Phi) + \frac{3}{2} \right)^{1/2}  \; ,
\end{equation}
where the relation between $\varphi$ and $\Phi$ depends on the exact form of the function $\omega(\phi)$. It brings the action in the following Einstein frame form already displayed in Eq.~\eqref{eq:bd_action_EinsteinFrame},
\begin{align}
\begin{split} \label{eq:bd_action_Einstein_app}
S &= \frac{1}{16 \pi} \int \mathrm{d}^4x \sqrt{-\tilde g} \left[ \tilde R -  \tilde g^{\mu \nu} \partial_\mu \varphi \partial_\nu \varphi \right] \\
& - \sum_A \int \mathrm{d} t \; m_A(e^{\Phi(\varphi)}) \sqrt{- e^{- \Phi(\varphi)}  \tilde g_{\mu \nu} v^\mu_A v^\nu_A} \; .
\end{split}
\end{align}
From now on we will drop the tildes for simplicity. The equations of motion for such an action are given in Eq.~\eqref{eq:system_ST}. However, in the main text the scalar charge defined in Eq.~\eqref{eq:def_scalar_charge} was a free parameter related to the short-distance physics coupling the scalar to the neutron star. In the PN literature, this charge can be related to the functions $\omega$ and $m_A$ in the weak-field approximation. This can be done by taking into account the point-particle source term in the scalar EOM. Thus, the field equation for $\varphi$ becomes
\begin{equation}
\frac{1}{\sqrt{-g}} \partial_\mu \left(\sqrt{-g} g^{\mu \nu}  \partial_\nu \varphi  \right) = 8 \pi \sqrt{-g_{00}} \frac{d\Phi}{d\varphi} \frac{d}{d\Phi} \left(m_A(e^\Phi) e^{-\Phi/2} \right) \delta^3(\mathbf{x}) \; ,
\end{equation}
where we recall that $\Phi$ is defined by eq. \eqref{eq:change_variable_ST}, and we have taken the point-particle to sit at the origin of the coordinates. In spherical coordinates, this is rewritten as
\begin{equation} \label{eq:field_eq_bd_charge}
\partial_r \left(r^2 a b  \bar \varphi'  \right) = 2 \frac{a}{\sqrt{\omega(e^\Phi) + 3/2}} \frac{d}{d\Phi} \left(m_A(e^\Phi) e^{-\Phi/2} \right) \delta(r) \; .
\end{equation}
The left-hand side contains the scalar charge of the body, while the right-hand side depends on the fields $a$ and $\Phi$ evaluated at the location of the point-particle (formally divergent in our approach where the cutoff scale corresponding to the inverse size of the neutron star is sent to infinity). We use a weak-field expansion in which $\Phi = \Phi_0 + \mathcal{O}(m/r)$, $a=1+\mathcal{O}(m/r)$ at the location of the point-particle (note that such expansion is not accurate for neutron stars where $m/r \sim \mathcal{O}(1)$: this is why we chose to consider the scalar charge as a free parameter in the main text). Integrating Eq. \eqref{eq:field_eq_bd_charge}, this yields
\begin{equation} \label{eq:def_scalar_charge_app}
\bar \varphi = \varphi_0 - \int \mathrm{d}r \; \frac{Q}{r^2 a b} \; ,
\end{equation}
where
\begin{equation}
Q = \sqrt{\frac{2 \zeta}{1 - \zeta}} m_A^{(0)} e^{-\Phi_0/2} (1-2s_A) \; ,
\end{equation}
where $\zeta$ is related to the asymptotic value of $\omega$ through Eq.~\eqref{eq:expansion_omega}, and the sensitivity $s_A$ is defined in Eq.~\eqref{eq:expansion_mA}. This is in agreement with the conventional PN literature \cite{Damour_1992, Damour:1993aa}. In the main text, we have made use of the reduced scalar charge $q = Q / m_A^{(0)}$.

Since the tadpole function $f$ was already given in Eq.~\eqref{eq:f_bd}, the only remaining task of this Appendix is to find the expansion of the mass function $\mu(r)$, defined in Eq.~\eqref{mu expansion}. From Eq.~\eqref{eq:bd_action_Einstein_app}, $\mu(r)$ is given by
\begin{equation}
\mu(r) = m_A\big(e^{\Phi(r)} \big) e^{- \Phi(r) / 2}
\end{equation}
Thus, we have to find the expansion of
 the function $\Phi$. 
By using that, from the very definition of $\Phi$ in Eq.~\eqref{eq:change_variable_ST},
\begin{equation} \label{eq:sensitivity}
\frac{\mathrm{d} \Phi}{\mathrm{d} r} = \frac{1}{\sqrt{\omega(e^\Phi) + 3/2}} \frac{\mathrm{d} \varphi}{\mathrm{d} r} \; ,
\end{equation}
one can parametrically solve for $\Phi$ in a $1/r$ expansion using the expansion of $\varphi$ given in Eq.~\eqref{eq:expansion_barphi}. This finally translates into the coefficients of $\mu(r)$,
\begin{align}
\begin{split}
\mu_0 &= m_A^{(0)} e^{-\Phi_0/2} \; , \\
\mu_1 &= \tilde q \left(s_A-\frac{1}{2} \right) \; , \\
\mu_2 &= \frac{\tilde q}{8 \zeta} \left( \zeta (4s_A - 2 + \tilde q(1-4s_A+4s_A^2+4s_A')) + 2\tilde q \omega_1(1-2s_A) \right) \; , \\
\mu_3 &= \frac{\tilde q}{48 \zeta^2} \left[ 8 \zeta ^2 (2 s_A-1)
 +6 \zeta  \tilde q \left(2 \omega_1+\zeta 
   \left(4 s_A^2-4 s_A+4 s_A'+1\right)-4 \omega_1
   s_A\right) \right. \\
    &+ \tilde q^2 \left\lbrace-\zeta  \left(2 \omega_1+24 \omega_1 s_A^2-16
   \omega_1 s_A+2 s_A+24 \omega_1 s_A'-1\right) \right. \\
   &+\left. \left. 2\zeta ^2 \left(4 s_A^3-6 s_A^2+12 s_A s_A'+4 s_A-6
   s_A'+4 s_A''-1\right)+4 (2 s_A-1) \left(4 \omega_1^2-\omega_2\right)\right\rbrace \right] \; ,
\end{split}
\end{align}
where $\tilde q = q \sqrt{\frac{\zeta}{2 -2\zeta}}$.

\section{Various functions and coefficients} \label{app:def_complicated_functions}

\subsection{Coefficients in the Lagrangian of odd modes}

In this Appendix we give the expressions of the functions $u_i$ and $\mathcal{G}_i$ respectively defined in the main text by Eq. \eqref{bulk action 1} and Eq. \eqref{q action}

\begin{subequations}
\begin{align}\label{u and v}
u_1 &= - \frac{\ell(\ell+1)}{4 a^4 c^3} \bigg[a c \left(a \left(b \left(\alpha  c^2 a''-3 a \alpha +a b M_{12}'-a (\alpha +2) c
   c''-a c \alpha '-2 a\right)-a (\alpha +2) c b'\right) \right. \nonumber \\
   & \left. +a c a' \left(\alpha  c b'+b
   \left(5 \alpha +2 b M_{12}'+c \alpha '+2\right)\right)+b c^2 \left(a'\right)^2
   \left(b M_{12}'-2 \alpha \right)\right) \nonumber \\
   &+2 a c M_{10}^2 \left(c \left(a \left(b c
   a''+a b'+a b c''\right)+a a' \left(c b'+3 b\right)-2 b c \left(a'\right)^2\right)-a^2
   b\right) \nonumber \\
   &+b M_{12} \left(c a'+a\right) \left(c \left(2 a \left(b c a''+a b'+a b
   c''\right)+a a' \left(2 c b'+3 b\right)-3 b c \left(a'\right)^2\right)-2 a^2
   b\right) \nonumber \\
   &+4 a^2 b c^2 M_{10} M_{10}' \left(c a'+a\right) - \frac{a^3 c}{b} (\ell - 1)(\ell+2) \bigg] \\
u_2  &= \frac{\ell (\ell+1) (\ell-1) (\ell+2) a b }{4 c^3}\left[2 b  c' M_{12}-c \left(1-2 M_{10}^2\right)\right]\, ,  \\
u_3  &= -\frac{\ell (\ell+1) b }{4
   a^2 c}\left[ \partial_r (a c) b M_{12} - a c \left(1-2 M_{10}^2\right)\right]\, ,  \\
u_4  &= \frac{1}{a b c
   \left[ \partial_r (a c) b M_{12} - a c \left(1-2 M_{10}^2\right)\right]}
   \bigg[ c^2 \left(b a'
   \left(b M_{12}
   a'+a \left(\alpha +2
   M_{10}^2-2
   \right)\right)+2
    a^2
   b'\right) \nonumber \\
   &+a b c c'
   \left(2 b
   M_{12}
   a'-a \left(\alpha -2
   M_{10}^2+2
   \right)\right)+a^2
   b^2 M_{12}
   c'^2 \bigg] \; .
\end{align}
\end{subequations}
The functions $\mathcal{G}_{00},\mathcal{G}_{rr}$ and $\mathcal{G}_{qq}$ that appear in the action \eqref{q action} are defined in terms of the above functions as follows~\cite{Franciolini:2018uyq}:
\begin{subequations}
\begin{align}
\mathcal{G}_{00} =& -\frac{u_3^2}{u_2}\, ,\\
 \mathcal{G}_{rr} =& \frac{u_3^2}{u_4 u_3' +u_3 \left(u_4'+u_4^2\right)-u_1}\, , \\
\mathcal{G}_{qq} =& \frac{(\mathcal{G}_{rr} )^2}{u_3^3} \left[u_3' \left(2 u_3' u_4'-u_1'\right)+u_3 \left(-u_4 u_1'-u_3'' u_4'+u_3' u_4''+u_4 u_3'
   u_4'\right) + u_1 u_3''\right .
   \\
  & \qquad \quad \qquad \qquad \qquad \left . +2 u_4 u_1 u_3'+u_1 u_3 \left(3 u_4'+u_4^2\right)-u_1^2+u_3^2 \left(u_4 u_4''-2
   u_4'^2\right)\right]\, . \nonumber 
\end{align}
\end{subequations}

\subsection{Power emitted} \label{power-app}
Here we report the expressions of the coefficients appearing in our main result, Eq.~\eqref{eq:power}:

\begin{align}
p_0 &= \frac{1}{36(1-2 \mu_1)^2}, \\
p_2 &= \frac{1}{504 (1-2 \mu_1)^4} \big( 23 + 112 a_2 - 140 \gamma_1 - 35 \lambda_0 - 260 \mu_1 + 
 280 \gamma_1 \mu_1 + 70 \lambda_0 \mu_1 + 92 \mu_1^2 \nonumber \\
 & + 
 28 \alpha_1 ( 2 \mu_1 - 1) + 224 \mu_2 \big), \\
p_4 &= \frac{1}{181440 (1-2 \mu_1)^6} \bigg( -33817+ 47040 \alpha_1^2 \mu_1^2-47040 \alpha_1^2 \mu_1+11760
   \alpha_1^2 \nonumber \\
   &-389760
   \alpha_1 \gamma_1 \mu_1+97440 \alpha_1 \gamma_1+97440 \alpha_1 \lambda_0 \mu_1^2-97440 \alpha_1
   \lambda_0 \mu_1+24360 \alpha_1 \lambda_0 \nonumber \\
   &+43200
   \alpha_1 \mu_1^3-427680 \alpha_1 \mu_1^2+483840
   \alpha_1 \mu_1 \mu_2+213840 \alpha_1 \mu_1-241920 \alpha_1 \mu_2 \nonumber \\
   &-5400 \alpha_1-80640 a_2^2+1440
   a_2 \big(84 \alpha_1 (2 \mu_1-1)+420 \gamma_1 (2
   \mu_1-1)+210 \lambda_0 \mu_1-105 \lambda_0 \nonumber \\
   &+184
   \mu_1^2+40 \mu_1-224 \mu_2+18\big)+60480 a_3 (2
   \mu_1-1)+80640 b_2 \mu_1^2-80640 b_2 \mu_1 \nonumber \\
   &+20160
   b_2+772800 \gamma_1^2 \mu_1^2-772800 \gamma_1^2 \mu_1+193200 \gamma_1^2+386400 \gamma_1 \lambda_0 \mu_1^2-386400 \gamma_1 \lambda_0 \mu_1 \nonumber \\
   &+96600 \gamma_1
   \lambda_0+1066560 \gamma_1 \mu_1^3-3414240 \gamma_1
   \mu_1^2+2419200 \gamma_1 \mu_1 \mu_2+1707120
   \gamma_1 \mu_1 \nonumber \\
   &-1209600 \gamma_1 \mu_2-133320
   \gamma_1-161280 \gamma_2 \mu_1^2+161280 \gamma_2
   \mu_1-40320 \gamma_2+48300 \lambda_0^2 \mu_1^2 \nonumber \\
   &-48300
   \lambda_0^2 \mu_1+12075 \lambda_0^2+283440 \lambda_0
   \mu_1^3-878760 \lambda_0 \mu_1^2+604800 \lambda_0
   \mu_1 \mu_2+439380 \lambda_0 \mu_1 \nonumber \\
   &-302400 \lambda_0 \mu_2-35430 \lambda_0-40320 \lambda_1 \mu_1^2+40320
   \lambda_1 \mu_1-10080 \lambda_1+117488 \mu_1^4 \nonumber \\
   &-182176
   \mu_1^3+449280 \mu_1^2 \mu_2-222648 \mu_1^2-167040
   \mu_1 \mu_2+241920 \mu_1 \mu_3+119096 \mu_1 \nonumber \\
   &-322560 \mu_2^2+213120 \mu_2-120960 \mu_3 +389760 \alpha_1 \gamma_1 \mu_1^2 \bigg).
\end{align}

In the simple case of a linear Gauss-Bonnet coupling discussed in Section \ref{sec:examples}, these expressions simplify however considerably. Using Eqs.~\eqref{eq:GB_coeffs}-\eqref{4-relation-charge}, we get that only $p_4$ deviates from its GR value,
\begin{align}
\begin{split}
p_0 &= \frac{1}{36} \; , \\
p_2 &= \frac{23}{504} \; ,  \\
p_4 &=  \frac{33817 + 236880 \frac{\bar \alpha^2}{M^4}}{181440} \; ,
\end{split}
\end{align}
where we recall that $\bar \alpha$ has been introduced in Eq. \eqref{eq:GB-action}.

\bibliographystyle{hunsrt}
\bibliography{test}

\end{document}